\documentclass{aa}
    
    \usepackage{graphicx}
    \usepackage[dvipsnames]{xcolor}
    \usepackage[colorlinks]{hyperref}
    \usepackage{txfonts}
    \usepackage{multirow}
    \usepackage{makecell}
    \usepackage{ulem}
    \usepackage{bm}
    \usepackage[dvipsnames]{xcolor}
    \usepackage{placeins}
    \usepackage{amsmath}

\begin{document}

    \title{Wolf-Rayet stellar evolution models with improved treatment of the atmosphere}
    
    \author{Thomas Voje\thanks{Corresponding author: thomas.voje@umontpellier.fr}
          \and Ana Palacios
          \and Fabrice Martins
          }

    \institute{LUPM, Univ. Montpellier, CNRS, Montpellier, France}

    \date{\today}

    \abstract
    {Evolutionary models of massive stars are quasi-exclusively computed using an Eddington gray atmosphere. This approximation does not accurately describe the complex physical phenomena occurring in the atmosphere of massive stars.}
    {We aim to include state-of-the-art atmosphere models in the evolution computations of massive stars and test how the Wolf-Rayet phase is impacted.}
    {We computed the evolution of Galactic massive stars with the code STAREVOL. During the advanced phases of evolution, we applied outer boundary conditions interpolated within a grid of CMFGEN model atmospheres at each time step. The effective temperature and effective gravity were extracted from the atmosphere models. We then compared the resulting evolutionary tracks with classical calculations assuming Eddington gray atmospheres.}
    {We find that including detailed model atmospheres has a significant impact on the effective radius and temperature of the models during the later stages of the evolution. The effective temperatures of the evolution models computed with detailed model atmospheres are greatly reduced and in better agreement with observations of Wolf-Rayet stars. On the other hand, the internal structure of the models is barely affected by the choice of the atmosphere. We show that applying post-processing corrections on effective temperature and gravity is a method equivalent to our direct inclusion of atmosphere models in evolutionary calculations.}
    {The inclusion of detailed atmosphere models in the computation of evolutionary models is necessary to correctly reproduce the position of evolved massive stars in the Hertzsprung-Russell diagram. However, this has no impact on the internal and chemical evolution.}


    \maketitle

\nolinenumbers

\section{Introduction}
\label{section:intro}

    Massive stars are of paramount importance in the evolution of the Universe. Despite their rarity and short lifetimes, they are major contributors to the production of heavy nuclei, which they expel into the interstellar medium through their strong winds along their evolution, when they explode as supernovae or through the merging of their compact remnants in binary systems \citep{Johnson_2019}. Being very luminous and hot stars, they also emit important quantities of UV radiations, largely contributing to the energetic balance of their host galaxies \citep{Eldridge_Stanway_2022}.    
    
    Evolutionary models are important numerical ingredients that are required to understand the quantitative role of massive stars in a galactic or cosmological context. Although successful in many ways, these models still suffer from limitations. In particular the effect of mass loss, especially for luminous blue variable (LBV) stars \citep{Owocki_et_al_2004,Smith_Owocki_2006,Cheng_et_al_2024,Pauli_et_al_2026}, the treatment of convection \citep{Jiang_et_al_2015,Agrawal_et_al_2022,Debnath_et_al_2024}, and the multiple effects of binarity \citep[see e.g.][]{Sana_et_al_2012,REVIEW_Marchant_Bodensteiner_2024} remain only partially understood.

    One of the remaining problems appearing when comparing stellar evolution models of massive stars to observations concerns the effective temperature of Wolf-Rayet (hereafter WR) stars. In general, the predicted effective temperatures associated with specific classes of WR stars (WN and WC stars) are higher than the ones deduced by the spectral analysis of stars using detailed atmosphere codes \citep{Hamann_et_al_2006,REVIEW_Crowther_2007,Hainich_et_al_2014,Hamann_et_al_2019}.\\ A possible explanation of this discrepancy could be the poor treatment of the atmosphere in stellar evolution models. Indeed, the physical phenomena arising in the atmosphere of massive stars are very complex, including but not limited to: non-local thermodynamic equilibrium (LTE) effects, non-gray effects, stellar winds, and line-blanketing \citep{Hubeny_Mihalas_2014}. These processes are taken into account in state-of-the-art radiative transfer codes such as CMFGEN \citep{CMFGEN_Hillier_Miller_1998}, PoWR \citep{PoWR_Grafener_et_al_2002}, or FASTWIND \citep{Santolaya-Rey_et_al_1997}, but not included in stellar evolution computations.

    In practice rigorous stellar atmosphere computations are too time-consuming to be performed in real time within stellar evolution codes. Consequently an approximate atmosphere calculated under Eddington gray conditions is routinely used as boundary conditions of the stellar structure equations.

    Some strategies have been developed in the past to try to overcome this approximate treatment of the atmosphere in evolutionary calculations.
    In order to account for the important optical thickness of WR winds, \cite{Schaller_et_al_1992} implemented a simple post-processing correction of the effective temperature during this evolutionary phase in the Geneva Evolution Code (GENEC) models. Under this framework there is no feedback from the atmosphere into the internal structure and evolution. The effective temperatures corrected in this way have since been incorporated in the model grids produced by the Geneva group \citep[see for instance][]{Ekstrom_et_al_2012}. However this method relies on a simplified model atmosphere that does not correctly represent the atmosphere of WR stars \citep{Groh_et_al_2014}.

    \cite{Groh_et_al_2014} improved the post-processing method by recomputing the effective temperature of preexisting stellar evolution computations using rigorous radiative transfer resolution models. In their study of a solar metallicity 60 $M_{\odot}$ model, the post-processed $T_{\rm eff}$ they obtained are compatible with the uncorrected ones on the main sequence evolution, but significantly lower during the WR phase, in better agreement with observations.
    A similar study by \cite{Josiek_et_al_2025} shows how the optical depth at which the model atmospheres are patched onto the precomputed evolution computation affects the post-processed effective temperature on the main sequence of a 150~$M_{\odot}$ stellar model.
    
    Finally, to bypass the issue of the unmanageable computational cost of coupling a stellar evolution code to a detailed radiative transfer model, \cite{Schaerer_et_al_1996} adopted another strategy. They coupled the Geneva code to a simplified atmosphere model computed with the ISAWIND code, including, for the first and only time to date, the feedback of the atmosphere onto the structure and evolution of massive star models. This method was applied to study the main sequence evolution of models with initial masses of 40, 60, and 85~$M_{\odot}$. They found that only their 85~$M_{\odot}$ model, close to the Eddington limit, was affected by this new treatment of the outer layers in the stellar evolution.

    Unlike what is now routinely performed for lower-mass stars
\citep{Morel_et_al_1994,Chabrier_Baraffe_1997,Montalban_et_al_2004,Mosumgaard_et_al_2018,Amard_et_al_2019},
    detailed atmospheres are not used in the computation of evolutionary models of massive stars.
    In this study we make a first step toward resolving this issue. We develop a method based on interpolations in precomputed grids of atmosphere models to provide boundary conditions in evolutionary calculations.
    In Sect.~\ref{section:tools} we describe the numerical tools we used to compute the evolution models and the model atmospheres. In Sect.~\ref{section:methods} we detail the method we used to include detailed model atmospheres in massive star evolution computations. In Sect.~\ref{section:results} we present the application of this method to compute nonrotating evolution models of WR stars with an initial mass of 50 and 60~$M_{\odot}$ at solar metallicity, along with the differences between such models and the same models computed using the Eddington gray approximation. We compare these results with other evolution models from the literature and with observations in Sect.~\ref{section:comparison}. We present some limitations of our models in Sect.~\ref{section:limitations}. Finally, we summarize the main conclusions of this study in Sect.~\ref{section:conclusion}.

\section{Numerical tools} \label{section:tools}

    In this section we describe the numerical tools we used to perform the computations. We computed the stellar evolution models with the version 3.40 of the stellar evolution code STAREVOL as in \citet[see also \citealt{Forestini_et_al_1991}, \citealt{Siess_et_al_2000}, and \citealt{Amard_et_al_2019} for details on the code]{Martins_Palacios_2017}.
    To compute the model atmospheres, we used the atmosphere code CMFGEN \citep{CMFGEN_Hillier_Miller_1998}.\\

    \subsection{General presentation of the STAREVOL code} \label{subsection:starevol_description}

    \paragraph{Resolution algorithm}
    STAREVOL is a one-dimensional stellar evolution code that solves the stellar structure and evolution equations expressed in the Lagrangian coordinate $m_r$, defined as the mass contained in a shell of radius $r$, using the Henyey method \citep{Henyey_et_al_1964}. In STAREVOL the stellar structure and evolution equations are solved from the center of the star (i.e., the first shell of the one-dimensional grid, where $m_r = 0$) to the numerical surface of the star (i.e., the last shell of the one dimensional grid, where $m_r = M_*$ with $M_*$ being the total mass of the star).

    \paragraph{Stellar structure and evolution equations}
    The stellar structure and evolution equations solved in STAREVOL are the following. \\
    
    \noindent Mass conservation equation:
    \begin{equation}
        \frac{\mathrm{d} r^3}{\mathrm{d} m_r} - \frac{3}{4 \pi \rho} = 0
        \label{eq:mass_conv}
    .\end{equation}
    Momentum conservation equation:
    \begin{equation}
        4 \pi r^2 \frac{\mathrm{d} P}{\mathrm{d} m_r} + \frac{G m_r}{r^2} + \frac{\mathrm{D}u}{\mathrm{D}t} = 0
    .\end{equation}
    Lagrangian velocity definition:
    \begin{equation}
        \frac{\mathrm{D}r}{\mathrm{D}t} - u = 0
        \label{eq:lagrangian_velocity}
    .\end{equation} 
    Energy conservation equation:
    \begin{equation}
        \frac{\mathrm{d} L}{\mathrm{d} m_r} - \sum_i \epsilon_i = 0
    .\end{equation}
    Energy transport equation:
    \begin{equation}
        \frac{\mathrm{d}  \ln T}{\mathrm{d} m_r} + \nabla \frac{1}{4 \pi r^2 P} \left(\frac{Gm_r}{r^2} + \frac{\mathrm{D}u}{\mathrm{D}t}\right)= 0
        \label{eq:energy_transf}
    .\end{equation}
The $\epsilon_i$ are the contributions by unit mass to the energy generation (gravitational contribution, $\epsilon_{\rm grav}$, contribution of nuclear fusion reactions, $\epsilon_{\rm nucl}$, and of the neutrinos, $\epsilon_{\rm neut}$, etc.), the temperature gradient, $\nabla = \mathrm{d} \ln T / \, \mathrm{d} \ln P$, is equal to the radiative gradient, $\nabla_{\rm rad}$, inside radiative zones and to the real gradient inside convective zones, and the other symbols have their usual meaning. To evolve the stellar models up to the part of the WR phase with approximately $T_{\rm eff} \ge 30$~kK, we assumed hydrostatic equilibrium so that $\mathrm{D}u/\mathrm{D}t = 0$ and Eq.~(\ref{eq:lagrangian_velocity}) was not taken into account. Beyond that point, the full set of stellar structure and evolution equations was solved with $\mathrm{D}u/\mathrm{D}t \neq 0$.

    \paragraph{Boundary conditions}
    To close the system, we need to apply boundary conditions at the center and at the numerical surface of the star. At the center the boundary conditions are naturally $r = 0$ and $L = 0$. At the numerical surface, the boundary conditions on $T$ and $\rho$ are given by a model atmosphere. This is the crucial part that is examined in detail in the present study. We describe the different model atmospheres that we used along with the associated boundary conditions in the following subsection.

    \paragraph{Mass loss}
    We used different mass loss prescriptions for the different spectroscopic phases that our evolution models go through. 
    The following recipes were adopted:
    \begin{itemize}
        \item For the OB phase with $\log(T_{\rm eff}) > 4$ and $X_{\rm H, surf.} > 0.4$, $X_{\rm H, surf.}$ being the surface mass-fraction of hydrogen, we used the prescription of \cite{Vink_et_al_2001}.
        \item For the LBV phase at $\log(T_{\rm eff}) < 4$, we used the \cite{deJager_et_al_1988} mass loss prescription, which we multiplied by a factor of 3 in order to account for eruptive mass loss events that we cannot model using regular mass loss prescriptions.
        \item In the WR phase where $\log(T_{\rm eff}) > 4$ and $X_{\rm H, surf.} < 0.4$ \citep{Meynet_Maeder_2003}, we used the \cite{Sander_Vink_2020} mass loss prescription in its $\Gamma_{\rm e}$ form.
        This prescription is based on hydrodynamical model atmospheres of WN stars and accounts for clumping.
        As this prescription was established using models with $X_{\rm H, surf.} = 0$, we added a surface hydrogen abundance dependency following \cite{Grafener_Hamann_2008}, and we also computed the Eddington factor used in the prescription as in \cite{Grafener_Hamann_2008}. The limitations of this extension of the \cite{Sander_Vink_2020} prescription to the WNh phase are discussed in Sect.~\ref{section:limitations}.
        
        As is described in Sect.~\ref{s_cmfgen}, iterations on the internal structure of the atmosphere models were performed in order to satisfy the momentum conservation equation. It turns out that in some regions of the parameter space explored in this study no solution could be obtained with this iterative process because of the proximity to the Eddington limit. As a consequence, we adopted a conservative approach and reduced further the mass loss rates of the evolutionary models by another factor of 3. This ensured that hydrodynamically consistent atmosphere structures could be computed for the entire grid of models.
        This additional reduction of the mass loss rate is also discussed in Sect.~\ref{section:limitations}.
        The final mass loss rate, $\dot{M}_{\rm WR}$, which we used for WR stars is thus
        \begin{equation}
        \begin{split}
            \log \dot{M}_{\rm WR} = & \: a\log\left[-\log(1-\Gamma_{\rm e})\right] - \log(2) \left(\frac{\Gamma_{\rm e,b}}{\Gamma_{\rm e}}\right)^c + d \\ & - 0.45 X_{\rm H, surf.} - 0.5
        \end{split}
        ,\end{equation}
        with $a = 2.932$, $\Gamma_{\rm e,b} = 0.244$, $c = 9.15$, $d = -2.61$ and
        \begin{equation}
            \Gamma_{\rm e} = 10^{-4.813} \frac{L}{M} (1+X_{\rm H,surf.}) 
        .\end{equation}
        \end{itemize}

    \paragraph{Convection}
    We determined the limits of convection zones using the Schwarzschild criterion for stability. We treated convection using the mixing length theory (MLT) with the formalism developed in \cite{Kippenhahn_et_al_1990} with a solar calibrated MLT parameter of $\alpha_{\rm MLT} = \Lambda / H_{\rm P} = 1.6602$, $H_{\rm P}$ being the pressure scale height. We also extended the size of the convective core by taking overshooting into account with a step overshooting parameter, $\alpha_{\rm ov} = d_{\rm ov}/H_{\rm P} = 0.15$.

    In the outer regions of massive stars, convection can become inefficient and lead to numerical difficulties. In order to bypass them, we chose to increase the $\alpha_{\rm MLT}$ parameter to an arbitrary value of 10 for the subsurface convection zones where $\log(T) < 6$. Other stellar evolution codes use similar ad hoc methods to compute the evolution of massive stars that become WR stars \citep{Maeder_1987,Alongi_et_al_1993,MESA_2_Paxton_et_al_2013,Agrawal_et_al_2022}.

    \paragraph{Reference chemical composition, opacities, and equation of state}
    We used the \cite{Asplund_et_al_2009} solar chemical composition with a calibrated solar metallicity of $Z_{\odot} = 0.013446$.
    We used for the opacities the OPAL tables \citep{OPAL_Iglesias_Rogers_1996}. For temperatures under $8\,000$ K, we used the opacities from \cite{Ferguson_et_al_2005}.
    We used the same equation of state as in \cite{Siess_et_al_2000}.

    \subsection{Atmosphere treatment in STAREVOL} \label{subsection:starevol_edd_gray}

    In the STAREVOL code, following the gray approximation, the optical depth is defined by $\mathrm{d}\tau = - \kappa \rho \: \mathrm{d}r$. The full set of stellar structure equations is solved from the center to the numerical surface that is set at an optical depth of $\tau = \tau_0$, with $\tau_0$ being one of the input parameters set to 0.005 by default. Hence, there is no decoupling in the treatment between the interior and the atmosphere as is the case in most of the other stellar evolution codes \citep{FRANEC,MESA_1_Paxton_et_al_2011,Cesam2k20}.

    \subsubsection{Eddington gray atmosphere: Classical models}

    The default setting of the code is to apply a plane-parallel Eddington gray model atmosphere as the upper boundary condition to the stellar structure equations. In the following, we refer to the resulting stellar evolution models as classical models. \\ In these classical models, the temperature in the atmosphere is a simple function of the optical depth, $\tau$:
    \begin{equation}
        T^4(\tau) = \frac{3}{4} T_{\rm eff}^4 \Bigl(\tau + q(\tau)\Bigl)
        \label{eq:Hopf}
    ,\end{equation}
    $T_{\rm eff}$ being the temperature of the equivalent black body, $T(\tau)$ the temperature profile, and $q(\tau)$ the Hopf function, which is constant and equal to $2/3$ in the Eddington gray approximation. It can, however, be generalized to correct the gray approximation, and other expressions exist \citep{Bohm-Vitense_1958,Henyey_et_al_1965,KrishnaSwamy_1966}. 
    
    In the Eddington gray case, the atmosphere is considered to comprise a single shell, corresponding to the numerical surface of the star. The boundary condition on the temperature is given by Eq.~(\ref{eq:Hopf}) and the boundary condition on the density is obtained from the momentum conservation. Both conditions apply at the numerical surface as follows:
    \begin{equation}
        T(\tau_0) = \left[\frac{3}{4} T_{\rm eff}^4 \left(\tau_0 + \frac{2}{3}\right)\right]^{1/4}
    ,\end{equation}
    \begin{equation}
        \rho(\tau_0) = \frac{\mu(\tau_0)}{\mathcal{R}T(\tau_0)}\left[\frac{\tau_0}{\kappa(\tau_0)}\left(\frac{Gm_r(\tau_0)}{r^2(\tau_0)} + \frac{\mathrm{D}u}{\mathrm{D}t}(\tau_0)\right) -\frac{\sigma}{c} T_{\rm eff}^4 \tau_0\right]
    .\end{equation}

    These two equations respectively represent the boundary conditions associated with Eqs.~(\ref{eq:energy_transf}) and (\ref{eq:mass_conv}).

    \subsubsection{Model atmospheres from grid interpolation: CMFGEN-based models} \label{subsection:realatmo}
    
    To account for detailed model atmospheres in low-mass evolution STAREVOL model computations, \cite{Amard_et_al_2019} used modified Hopf functions computed at each time step. They relied on interpolations within a grid of PHOENIX model atmospheres. \\ For the evolved massive star models considered in the present study, we adopted another method and instead “pasted” a model atmosphere -- i.e., temperature, density, and radius profiles -- interpolated within a grid of CMFGEN model atmospheres
    (see Sect.~\ref{subsection:cmfgen}) to the internal structure equations. In the following we refer to these new models as CMFGEN-based models.
    
    Let $T_{\rm atm}(\tau)$, $\rho_{\rm atm}(\tau)$, and $r_{\rm atm}(\tau)$ be, respectively, the interpolated temperature, density, and radius atmosphere profiles. We matched the temperature and density profiles in the deep regions of the atmosphere. The numerical surface of the star was set at $\tau_0 = 10$, and we replaced the Eddington gray surface boundary conditions at the numerical surface of the stellar model by the atmosphere temperature and density profiles values $T_{\rm atm}(\tau_0)$ and $\rho_{\rm atm}(\tau_0)$. We thus fed feedback from the interpolated detailed model atmospheres into the evolution model. The method is fully described in Sect.~\ref{section:methods}.

    \subsection{Stellar atmosphere code CMFGEN} \label{subsection:cmfgen}
    \label{s_cmfgen}

     Model atmospheres were computed with the code CMFGEN \citep{CMFGEN_Hillier_Miller_1998}. It is widely used for the analysis of various types of hot massive stars \citep[e.g.][]{hil03,martins04,bouret13,aadland22,martins24}. CMFGEN solves the statistical equilibrium and radiative transfer equations under non-LTE conditions. Sphericity is assumed and a stellar wind is included. A quasi-hydrostatic photosphere is connected to an expanding atmosphere parameterized by a velocity law of the form $v = v_{\infty} \times (1-R/r)$, where $R$ is the stellar radius at the photosphere, $r$ the radial coordinate, and $v_{\infty}$ the maximum velocity at the top of the atmosphere, i.e., at a radius equal to 100 times the stellar radius. The inner quasi-hydrostatic solution is iterated to take into account the radiative acceleration caused by line absorption, and thus provides a consistent velocity structure in the inner atmosphere. 

     The input parameters of the model atmospheres are the stellar luminosity, surface gravity, stellar radius, mass loss rate, terminal velocity, $v_{\infty}$, and the elements (and their various ions) to be taken into account at a given chemical composition. Effective temperature and mass are also given for consistency checks with luminosity, radius, and gravity. Continuum and line opacities are computed using atomic data and atmospheric conditions. Thousands of atomic levels and lines are taken into account. Iterations of the radiative transfer and statistical equations are performed until convergence is achieved. Convergence is evaluated on 1) the relative changes of atomic level populations from one iteration to the next and 2) the conservation of luminosity throughout the atmosphere.  

     A final model atmosphere consists of a density and temperature profile together with the optical depth scale along the radial coordinate. In the following we describe how we used CMFGEN to provide detailed model atmospheres for evolutionary calculations.

\section{Method} \label{section:methods}

    In this section we describe the method we developed to include detailed model atmospheres in stellar evolution computations for massive stars. Previous studies have shown that including detailed atmospheres in evolution models has a negligible impact on the main sequence evolution of solar metallicity massive stars \citep{Schaerer_et_al_1996, Groh_et_al_2014}. We thus computed the evolution using a classical (Eddington gray atmosphere) setup up to the beginning of the WR phase where we applied the method described hereafter. The strategy we adopted was to interpolate atmosphere profiles within a precomputed grid of CMFGEN models and to use these profiles as boundary conditions in the evolution computations.

    \subsection{Grid of model atmospheres} \label{subsection:grid}

    Here we describe the grid of CMFGEN model atmospheres together with the interpolation parameters and interpolation method. We first built a grid of CMFGEN model atmospheres encompassing the classical tracks in the relevant parameter space. We then performed the evolutionary calculation using detailed atmospheres, and extended the grid of atmosphere models when needed so that the final grid encompasses the parameter space of the new evolutionary tracks. The final grid is depicted in Fig.~\ref{fig:grid_60Msun} for our 60~$M_{\odot}$ track\footnote{The grid for the 50~$M_{\odot}$ model is shown in Appendix~\ref{ap_1}.}. In particular, the models of this grid take into account the evolution of the surface abundances along the track as predicted from stellar evolution models. This grid density ensures the smooth operation of the interpolation method (see Sect.~\ref{subsection:delaunay}).

  \begin{figure}
        \centering
        \includegraphics[width=0.95\hsize]{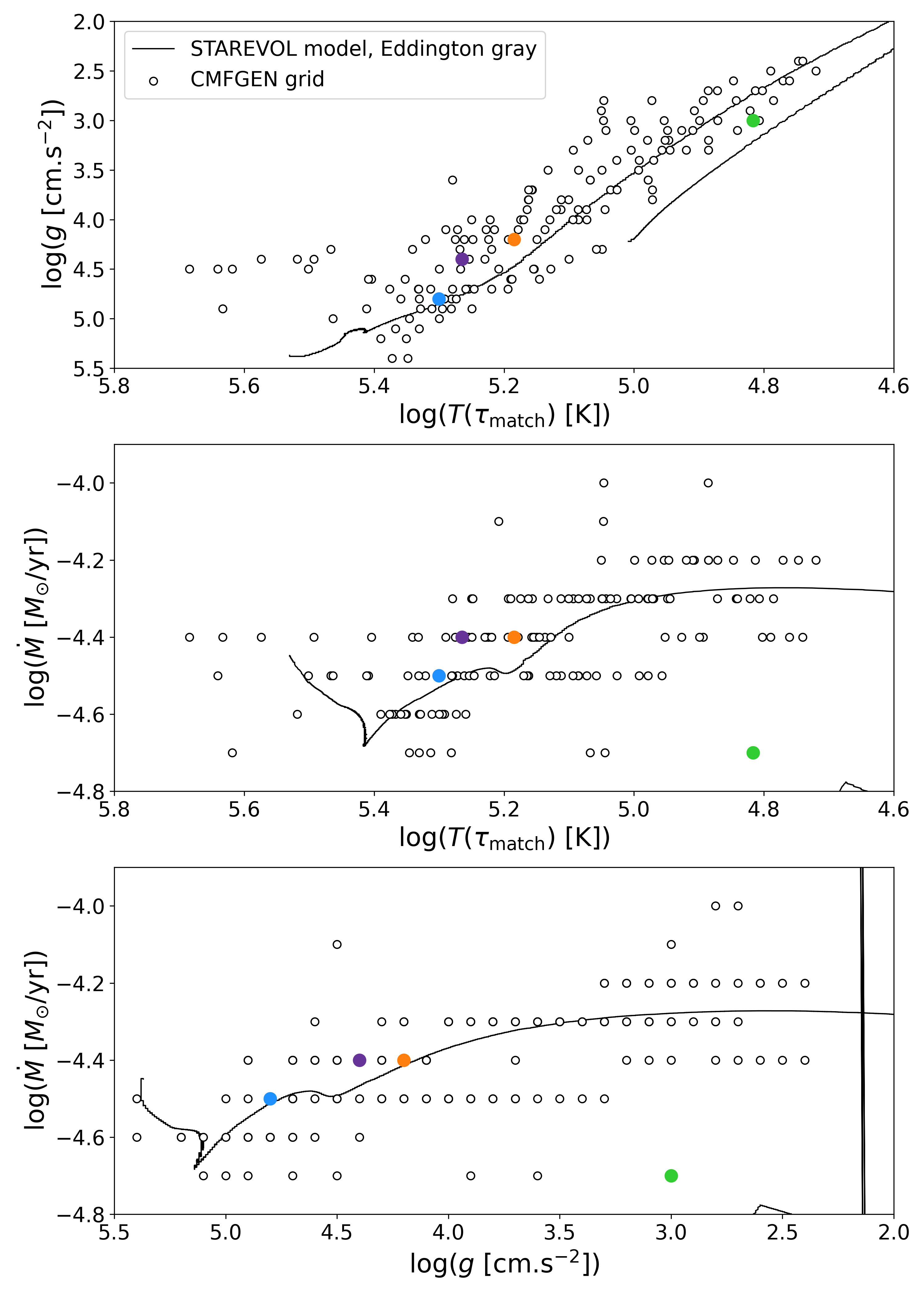}
        \caption{Grid of CMFGEN models in the $\log(g)$ vs. $\log(T(\tau_{\rm match}))$ diagram (upper plot), in the $\log(\dot{M})$ vs. $\log(T(\tau_{\rm match}))$ diagram (middle plot) and in the $\log(\dot{M})$ vs. $\log(g)$ diagram (lower plot), along with the evolutionary track of the classical 60 $M_{\odot}$ stellar model we use. $T(\tau_{\rm match})$ is the temperature at the matching point set at $\tau_{\rm match} = 25$ (see text). The green, orange, purple, and blue dots designate the points at which CMFGEN profiles are presented in Fig.~\ref{fig:parameter_impact}.}
        \label{fig:grid_60Msun}
    \end{figure}

    \begin{figure*}[ht]
        \centering
        \includegraphics[width=0.9\hsize]{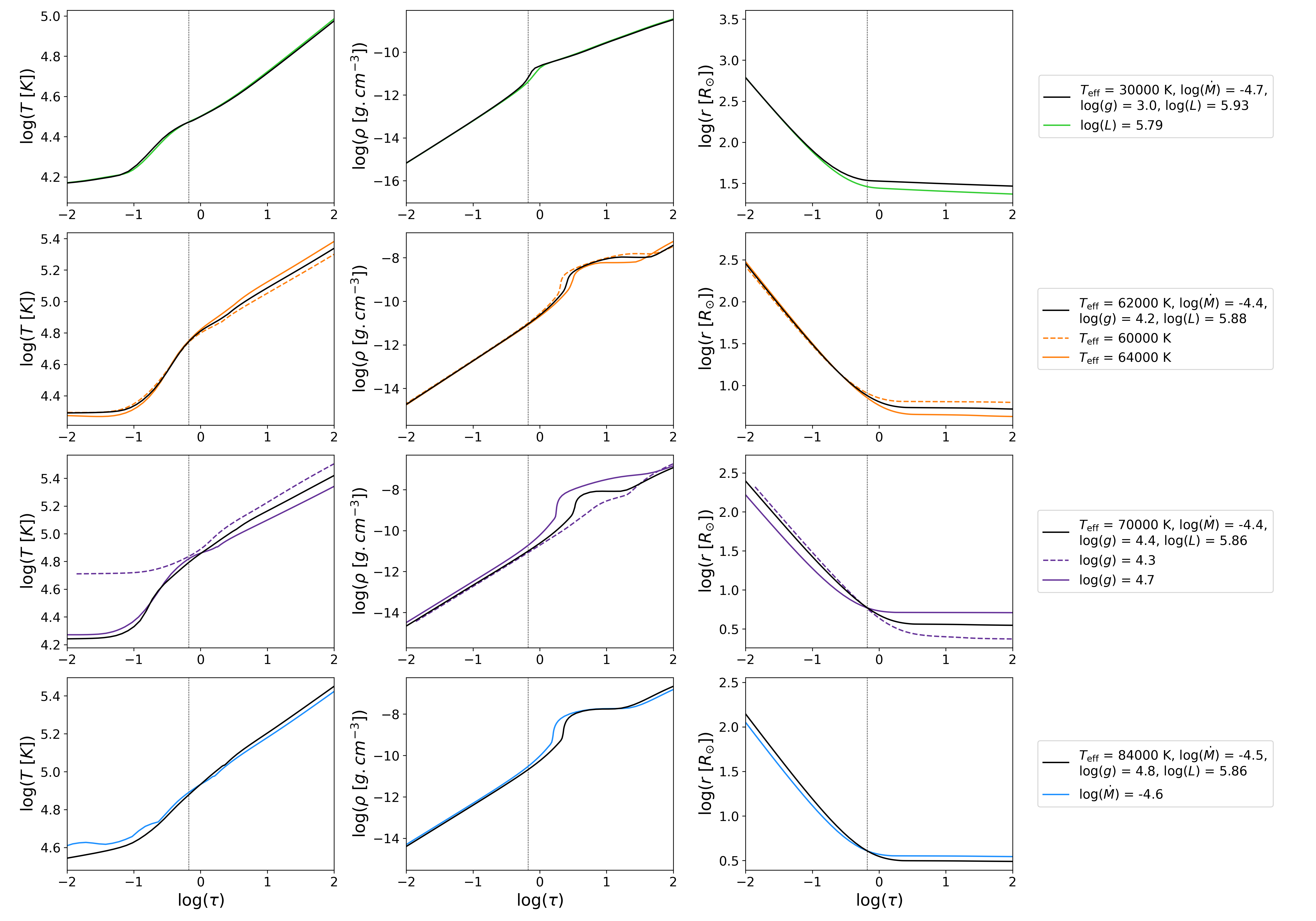}
        \caption{Impact of individual variations in the luminosity (first row), effective temperature (second row), surface gravity (third row), and mass loss rate (fourth row) on CMFGEN temperature (left column), density (central column), and radius (right column) profiles. Colored profiles correspond to CMFGEN models computed with the same stellar parameters as the black profile except for the parameter indicated in the legend. The luminosity is expressed in solar units, the mass loss rate in solar masses per year, and the gravity in centimeters per second squared. Each row refers to a grid point indicated by a dot of the corresponding color in Fig.~\ref{fig:grid_60Msun}. The vertical dotted lines indicate the photosphere ($\tau_{\rm eff} = 2/3$).}
        \label{fig:parameter_impact}
    \end{figure*}

    \subsubsection{Interpolation parameter space}

    In order to determine the interpolation parameters to be considered, we first studied the impact of input stellar parameters on CMFGEN atmosphere profiles.
    The relevant input parameters of the model atmospheres are: the luminosity ($L$); the effective temperature ($T_{\rm eff}$), which together with the luminosity constrains the photospheric radius;
the gravity at the photosphere ($g$); the mass loss rate ($\dot{M}$); and the surface chemical mass fractions.
    The surface abundances were not used in the interpolation method as they were self-consistently treated in the model atmospheres according to the effective temperature, gravity, luminosity, and mass loss rate of the corresponding stellar evolution model. It is thus sufficient to interpolate within these parameters. The terminal velocity was estimated from the escape velocity, with a classical scaling factor of 2.6 between both of them \citep[e.g.][and references therein]{Garcia_et_al_2014}.

    In Fig.~\ref{fig:parameter_impact} we show the impact of varying the luminosity, effective temperature, surface gravity, and mass loss rate, while keeping the other parameters constant, on the temperature, density, and radius profiles in CMFGEN model atmospheres.
    The luminosity has a very small impact on the temperature and density profiles, but strongly impacts the radius profile. As the evolutionary path of our models on the WR phase is at an almost constant luminosity
    ($\Delta \log(L/L_{\odot}) < 0.15$ over the entire evolution) and the computational cost of the CMFGEN models is expensive, we did not take the luminosity into account in our interpolation. Instead
    we made sure that the luminosity of the CMFGEN models of our grid remained close to
    the one of the evolutionary track and we corrected the radius profile after the interpolation was performed (see Sect.~\ref{subsection:delaunay} for details on this correction). On the other hand, effective temperature, mass loss rate and surface gravity have an impact on temperature and density profiles in the deep atmosphere (for $\tau \ge 10$), where the connection was performed. \\

    At large optical depth, in the region where the diffusion approximation is valid, the physics of radiative energy transport is the same in STAREVOL and CMFGEN models. Therefore the interpolated atmosphere temperature profile should match the stellar interior temperature profile.
    We thus chose to match $T_{\rm atm}(\tau)$ to $T(\tau)$ at $\tau = \tau_{\rm match} = 25$ 
    \footnote{We also checked that using a deeper matching point at $\tau_{\rm match} = 50$ does not change the results.}. We determined~$T(\tau_{\rm match})$ using cubic spline interpolation \citep{Press_et_al_1992} in both the interior and the atmosphere, and used it instead of $T_{\rm eff}$ as an interpolation parameter within the grid of model atmospheres alongside $\log(g)$ and $\log(\dot{M})$.

\begin{figure}[t]
        \centering
        \includegraphics[width=0.95\hsize]{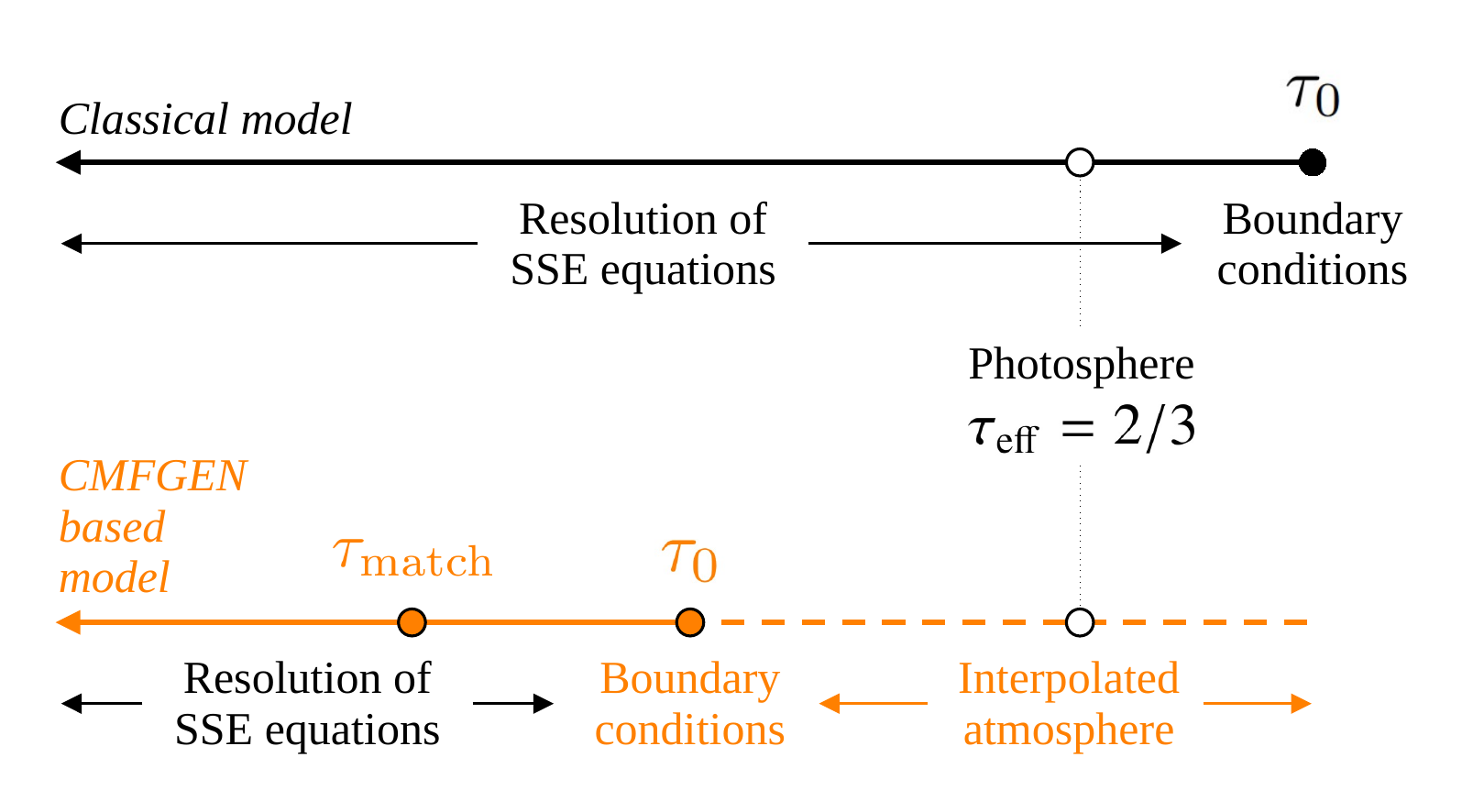}
        \caption{Sketch of the structure of a classical model (upper part - black) and a CMFGEN-based model (lower part - orange) as a function of the optical depth, $\tau$. $\tau_0$ is the numerical surface of the evolution model, $\tau_{\rm match}$ is the optical depth at which the interpolated model atmosphere temperature profile, $T_{\rm atm}(\tau)$, matches the internal structure temperature profile, $T(\tau)$ (see Sect.~\ref{subsection:grid}), and $\tau_{\rm eff} = 2/3$ defines the photosphere.}
        \label{fig:sketch_struc}
    \end{figure}

    Before performing any interpolation, we normalized the interpolation parameters $T(\tau_{\rm match})$, $\log(g)$ and $\log(\dot{M})$ with respect to the size of the grid so that the interpolation parameter space only contained points located inside the unit cube. For each model, $n$, we then remapped the CMFGEN's $T_{\mathrm{atm,}n}(\tau)$, $\rho_{\mathrm{atm,}n}(\tau)$ and $r_{\mathrm{atm,}n}(\tau)$ onto the optical depth grid of the STAREVOL structure for $\tau \ge \tau_0$ using cubic spline interpolations. For $\tau < \tau_0$ we made sure to always have a point at $\tau_{\rm eff} = 2/3$ in the interpolated atmospheric profiles, which is where we define the photosphere (see Fig.~\ref{fig:sketch_struc}).

    \subsubsection{Delaunay triangulation based interpolation method}\label{subsection:delaunay}
    
    As the grid of CMFGEN models is sparse and not cartesian, we performed the interpolation using a Delaunay triangulation based method. The Delaunay triangulation is an unstructured simplicial mesh that can be used for multivariate interpolation. More specifically, we adopted the Fortran algorithm developed under the name {\sc delaunaysparse} by \cite{DELAUNAYSPARSE_Chang_et_al_2018,DELAUNAYSPARSE_Chang_et_al_2020} that allows a gain in computational time for interpolation in medium to high dimensions.

    The Delaunay triangulation of the CMFGEN grid is first computed in the normalized stellar parameter space. Let $\bm{p}$ be the point in the normalized stellar parameter space we want to interpolate at. 
    For this interpolation method to be applicable, this point must be located inside one of the simplices of the Delaunay triangulation (here the simplices are tetrahedra, since we are in a three-dimensional parameter space).
    When building the grid, we thus made sure that almost every point of the evolutionary track was located inside one of the simplices of the grid. When this was not the case, we used the extrapolation method described in \cite{DELAUNAYSPARSE_Chang_et_al_2020}.

    A sketch of a Delaunay simplex is given in Fig.~\ref{fig:delaunay}. We computed the barycentric coordinates, $w_n$, of the interpolation point, $\bm{p}$, such that $\bm{p} = \sum_n w_n \bm{s_n}$, $\sum_n w_n = 1$, and $w_n > 0$ for $n = 1,2,3,4$. The interpolated temperature, density, and radius, $T_{\rm atm}$, $\rho_{\rm atm}$, and $r_{\rm atm}$, at a given optical depth, $\tau$, were then computed as a linear combination of the values at the vertices, with the barycentric coordinates of the interpolation point being used as the associated weights:

        \begin{figure}[t]
        \centering
        \includegraphics[width=0.9\hsize]{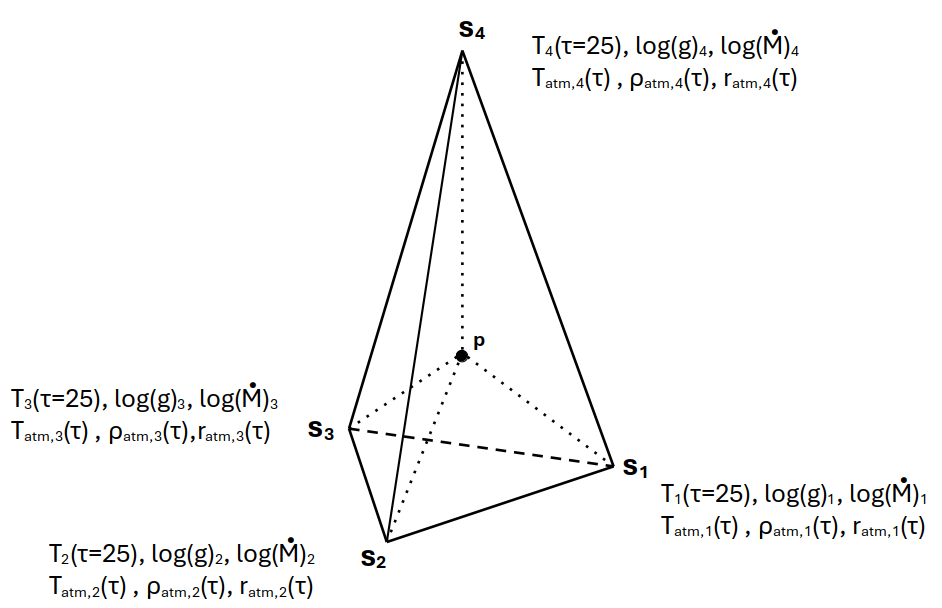}
        \caption{Representation of a point in a tetrahedral simplex of the Delaunay triangulation. $\bm{p}$ is the point in the normalized stellar parameter space we want to interpolate at and the $\bm{s_n}$ are the vertices of the simplex the point, $\bm{p}$, is in ($n = 1,2,3,4$).}
        \label{fig:delaunay}
    \end{figure}

    \begin{equation}
        T_{\rm atm}(\tau) = \sum_n w_n \, T_{\mathrm{atm,}n}(\tau)
    ,\end{equation}
    \begin{equation}
        \rho_{\rm atm}(\tau) = \sum_n w_n \, \rho_{\mathrm{atm,}n}(\tau) 
    ,\end{equation}
    \begin{equation}
        r_{\rm atm}(\tau) = \sum_n w_n \, r_{\mathrm{atm,}n}(\tau) 
    .\end{equation}

    As the luminosity of the model atmospheres used for the interpolation was not the same as the one from the evolution model, we scaled the interpolated radius profile, $r_{\rm atm}$, by a factor of $\sqrt{L/L_{\rm atm}}$, where $L$ is the total luminosity of the evolution model and $L_{\rm atm} = \sum_n w_n \, L_{\mathrm{atm,}n}$ is the luminosity of the interpolated model atmosphere.

\begin{figure}[t]
        \centering
        \includegraphics[width=0.9\hsize]{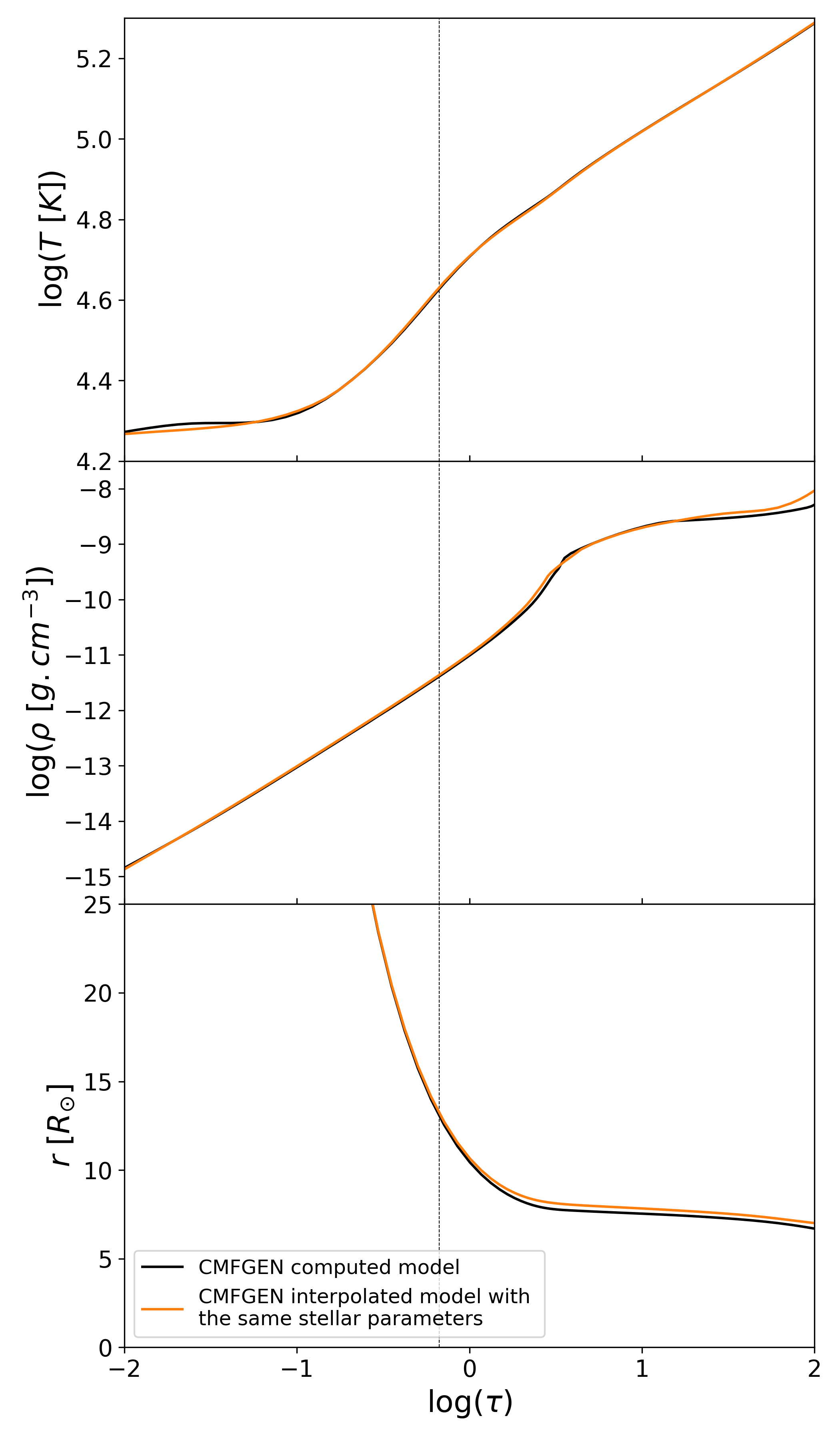}
        \caption{Temperature, density, and radius profiles produced by interpolating a CMFGEN model inside the grid using {\sc delaunaysparse} (orange) with the same interpolation parameters ($T(\tau_{\rm match}) = 132\,865$~K, $\log(g) = 3.638$, $\log(\dot{M}) = -4.328$), and luminosity ($\log(L/L_{\odot}) = 5.935$) as a WR model computed with CMFGEN (black).}
        \label{fig:idw_check}
    \end{figure}

    We checked that this method is able to correctly reproduce CMFGEN's temperature, density, and radius profiles by comparing interpolated profiles to profiles computed directly by CMFGEN with the same set of input parameters. Figure~\ref{fig:idw_check} presents the result of such a comparison. The interpolated profiles reproduce the actual profiles with adequate accuracy, which validates the interpolation method.

    \subsection{Inclusion of atmospheres in stellar evolution models}
    
\begin{figure}
        \centering
        \includegraphics[width=0.8\hsize]{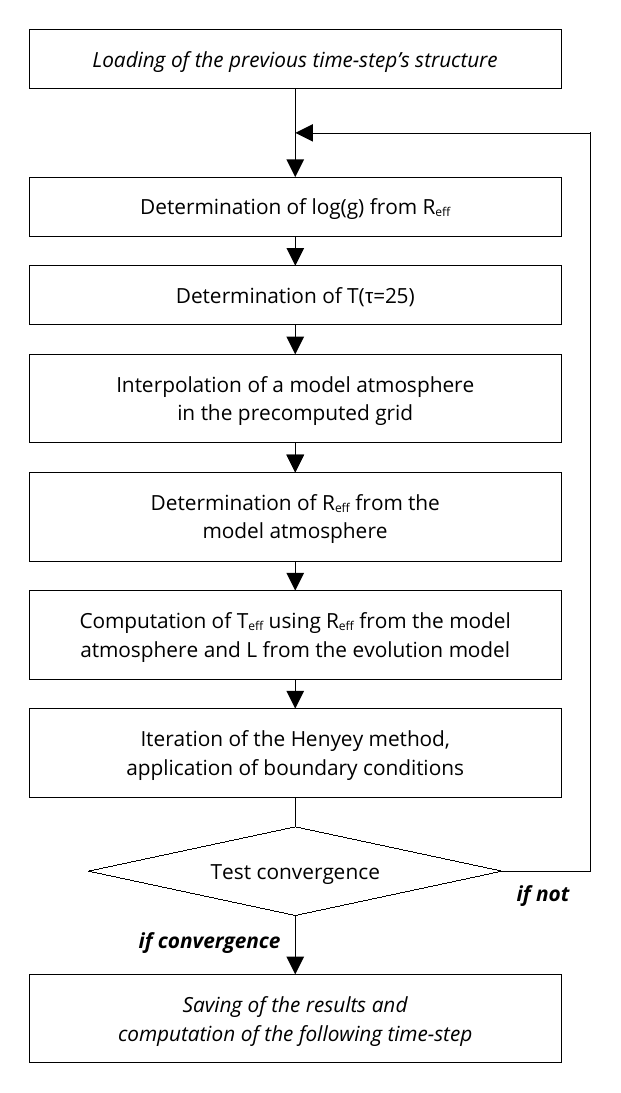}
        \caption{Flowchart of the inclusion of the CMFGEN model atmospheres in STAREVOL.}
        \label{fig:flowchart}
    \end{figure}

    A summary flowchart of the method is given 
    in Fig.~\ref{fig:flowchart}. The general process is the following:
    (1) for a given time step within the stellar evolution code, use the value of $T(\tau_{\rm match})$, $\log(g)$, and $\log(\dot{M})$ from the previous iteration step during stellar structure resolution\footnote{For the first iteration, the values from the previous time step are used.} and interpolate at this point the temperature, density, and radius profiles among the atmosphere models grid (see Sect.~\ref{subsection:grid}); (2) use the interpolated atmospheric temperature and density values at $\tau_0 = 10$ as boundary conditions to the interior stellar structure (see Sect.~\ref{subsection:realatmo}); (3) use the interpolated atmospheric radius profile to determine the photospheric radius, $R_{\rm eff}$, and relevant photospheric quantities ($T_{\rm eff}$ and $g$); (4) repeat until convergence of the Henyey scheme for the internal structure; (5) store the photospheric quantities from the last interpolated atmospheric structure; (6) move to the next time step and repeat the procedure from (1).

    In this procedure, the resulting stellar structures were divided into two parts as illustrated by the sketch in Fig.~\ref{fig:sketch_struc}:
    \begin{itemize}
        \item for $\tau > \tau_0$, we used the internal structure computed with STAREVOL;
        \item at $\tau = \tau_0$, we used the value of 
        the temperature and density
        from the interpolated model atmosphere profiles as boundary conditions to the internal structure equations;
        \item for $\tau < \tau_0$, we used the interpolated atmosphere structure, considering that the mass, the luminosity, and the chemical composition do not vary in this region.
    \end{itemize}
    The numerical surface of the interior model was set at $\tau_0 = 10$, much larger than the value used for a classical model, which was set at 0.005.
    This was done so that the numerical surface was located in a region where the diffusion approximation is valid and the Rosseland opacities can be considered to be equal to the flux-weighted opacities.

\section{Results} \label{section:results}

    We now describe how our revised treatment of the inclusion of atmosphere models in the computation of evolutionary models affects the internal structure and surface properties of evolved massive stars. We focus on the advanced phases of the evolution of 50~$M_{\odot}$ and 60~$M_{\odot}$ stars. These masses were chosen to allow for comparisons with known Galactic WR stars (Sect.~\ref{section:comparison}). The masses of these models when we start including detailed atmospheres are 20~$M_{\odot}$ and 29~$M_{\odot}$, respectively.
    
    In this section we illustrate our results using our 60~$M_{\odot}$ evolutionary models. These results are also valid for our 50~$M_{\odot}$ models.

    \subsection{Impact of detailed atmosphere on stellar structure}
    
    \begin{figure}[t]
        \centering
        \includegraphics[width=0.9\hsize]{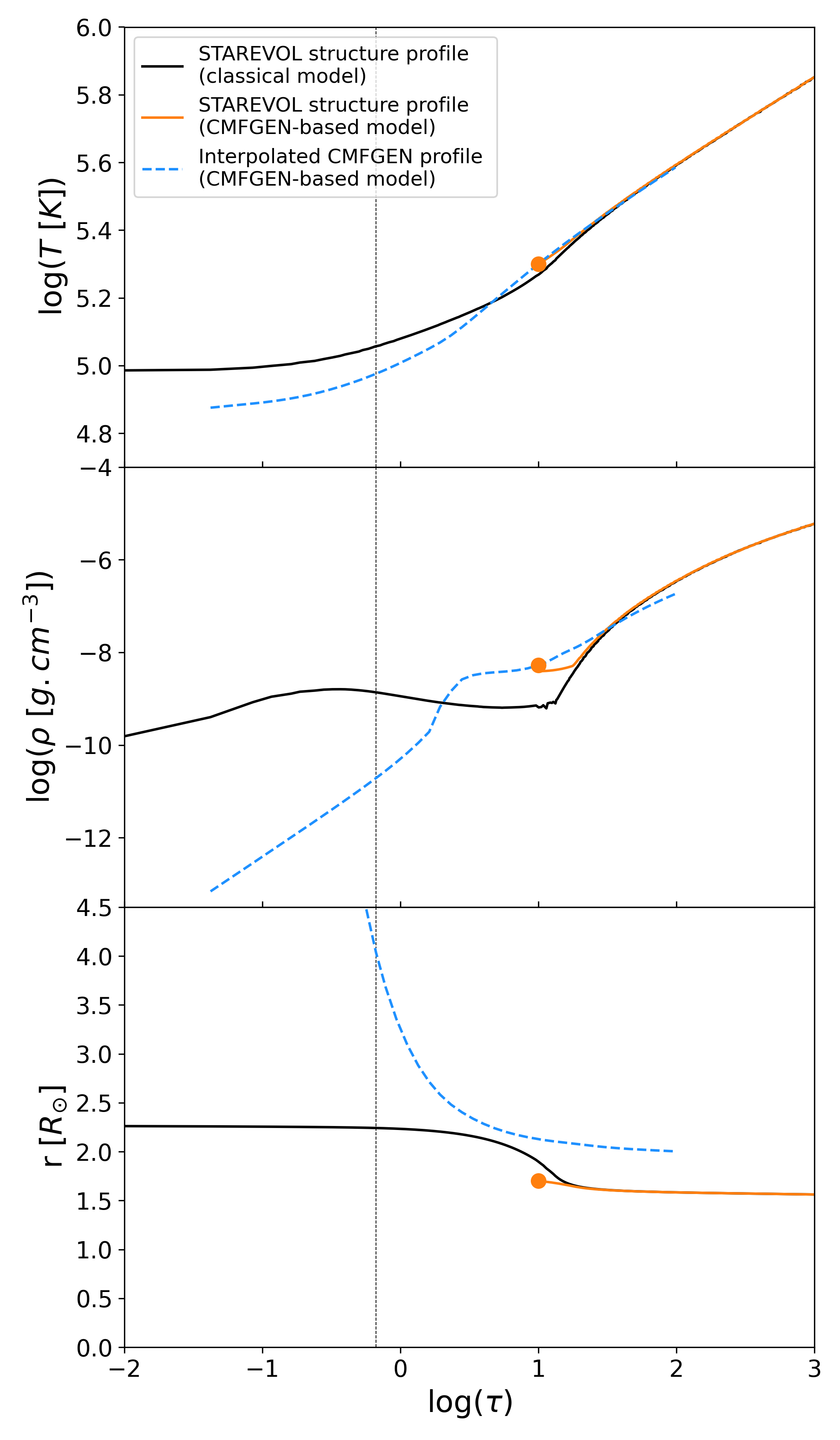}
        \caption{Temperature, density, and radius as functions of the optical depth for the classical 60~$M_{\odot}$ model (black), and the CMFGEN-based 60~$M_{\odot}$ model (orange: interior profile, blue: interpolated atmospheric profile). The vertical dotted line indicates the photosphere ($\tau_{\rm eff} = 2/3$) and the location of $\tau_0$ is indicated by a point. All the profiles correspond to a point in the evolution where the luminosity is $\log(L/L_{\odot}) = 5.891$ and marked on the evolutionary tracks in Fig.~\ref{fig:hrd_WR}. The effective temperature of the classical model at this point is $T_{\rm eff} = 114\,612$~K.}
        \label{fig:profiles_WR}
    \end{figure}

    We show in Fig.~\ref{fig:profiles_WR} the temperature, density, and radius profiles for the points marked by dots in the Hertzsprung-Russell diagram of the 60 $M_\odot$ models shown in Fig.~\ref{fig:hrd_WR}.
    The CMFGEN temperature profile correctly matches the internal profile
    in the inner region of the atmosphere (i.e. at $\tau = \tau_0 = 10$). Furthermore the temperature gradient is also continuous, which is expected because the diffusion approximation is valid in this region, and thus the physics of radiative energy transport is the same in both the evolution model and the model atmosphere. In the outer layers, the two profiles radically diverge.
    The density profile in the CMFGEN model atmosphere differs from that of the classical model, particularly at a small optical depth. The atmosphere density profile presents an inflection point around $\tau \approx 3$ that indicates the region from which the stellar wind is launched. The connection between the interior density profile and the associated atmosphere at $\tau=10$ triggers a bending of the inner density profile to higher values of $\rho$ compared to the classical model due to the lever arm of the boundary condition. This boundary condition ensures a continuity of the density profile but not of the density gradient.
    
    As the structure equations solved in the atmosphere and the evolution codes are not the same, the density profiles have very different shapes. This implies that the radius profiles are discontinuous.
    This discontinuity has no impact on the internal stellar structure, since the atmospheric radius profile is not used to provide boundary conditions to the evolution model (see Sect.~\ref{section:tools}).
    The atmospheric radius profile is, however, fully consistent with the temperature and density profiles, which apply feedback to the structure, and can thus be used to set the photospheric radius. It is interesting to note that the internal radius is smaller in the CMFGEN-based model than in the classical model in the region around $\tau = 10$ due to a larger density there.
    As was stated before, the onset of the wind is commonly located below the photosphere for WR stars \citep{REVIEW_Crowther_2007}. We thus observe a clear expansion of the atmosphere in the outer layers due to the stellar wind. This is not seen in the classical model, which implements a static plane-parallel atmosphere with no wind. The expanded atmospheric radius translates into a photospheric radius about 80\% larger compared to the classical model for the time step depicted in Fig.~\ref{fig:profiles_WR}, and thus yields an effective temperature that is lower by more than 29~kK for the same luminosity when the detailed model atmosphere is taken into account.
    
    We see that this feedback impacts the outer regions of the evolution model.
    On the other hand, we show that the impact of this feedback on the internal structure of the stellar model is negligible, as all profiles clearly match the classical model at large optical depths ($\tau \ge 100$).

    \subsection{Impact of detailed atmosphere on stellar evolution} \label{subsection:evolution}

    \begin{figure*}[t]
        \centering
        \includegraphics[width=0.49\hsize]{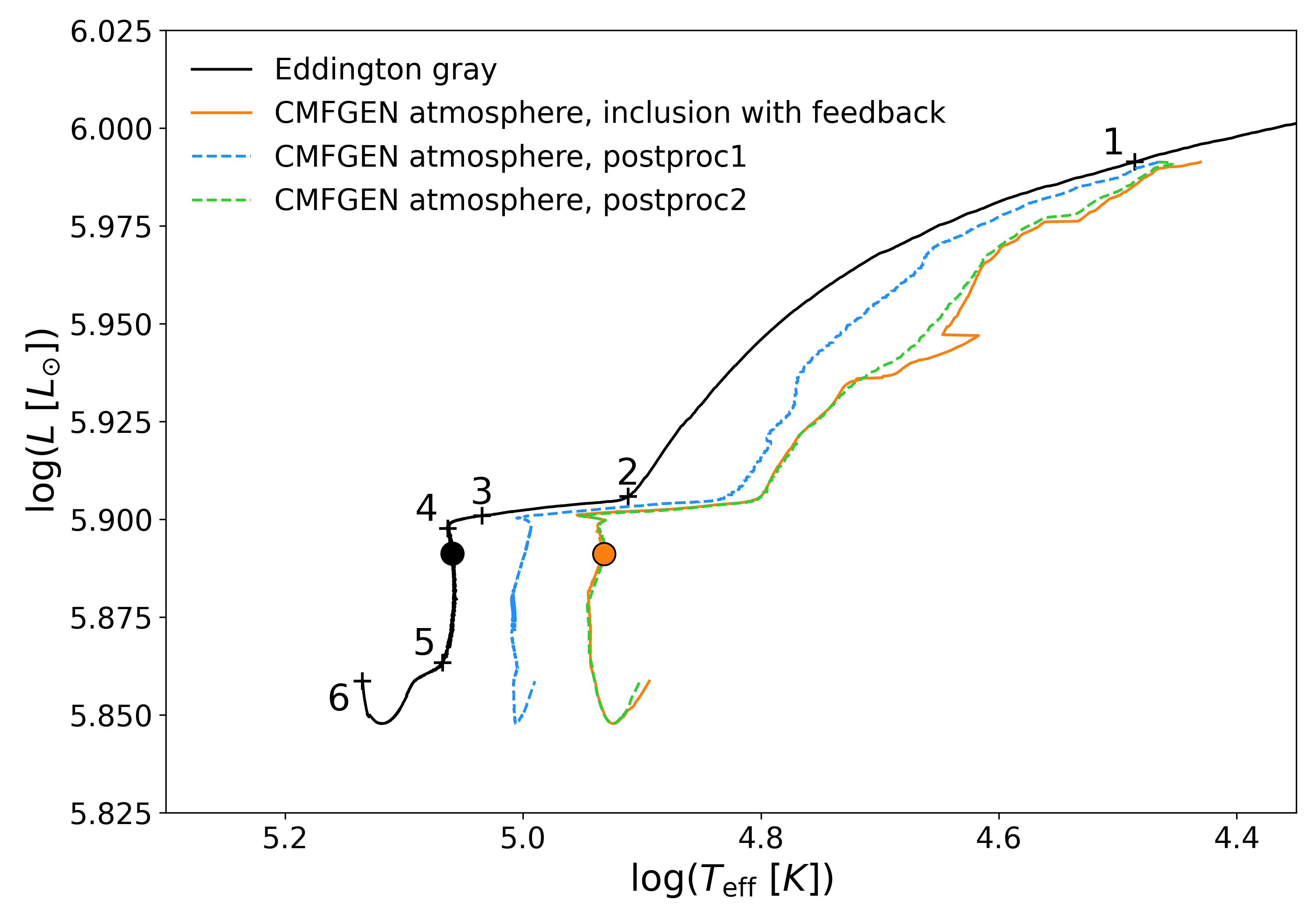}
        \includegraphics[width=0.49\hsize]{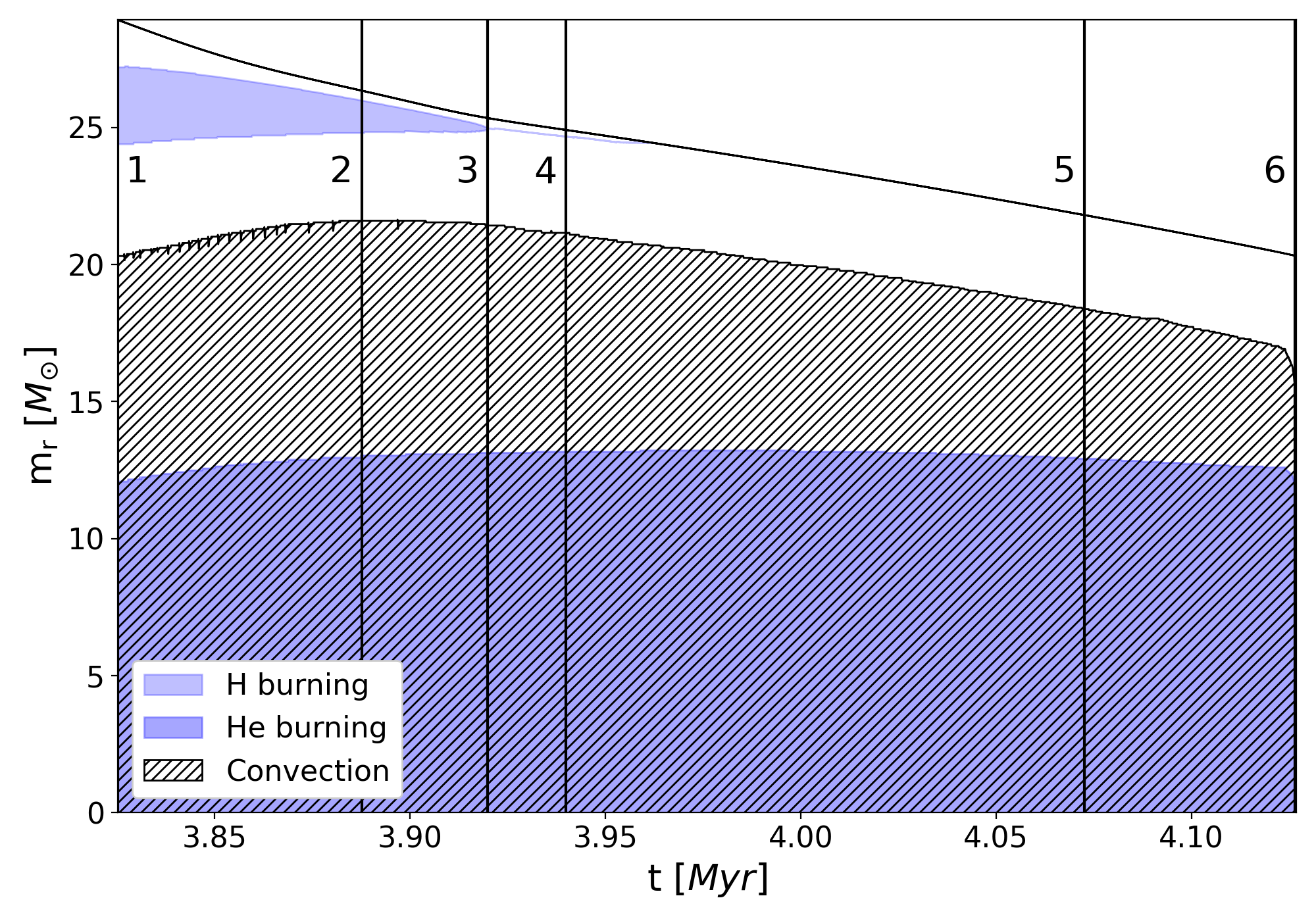}
        \caption{\textit{Left}: Evolutionary tracks of four different 60~$M_\odot$ nonrotating models at solar metallicity in the Hertzsprung-Russell (HR) diagram. The black line corresponds to the classical model, the orange line corresponds to the CMFGEN-based model as described in Sect.~\ref{section:methods}, and the blue line and the green line correspond, respectively, to the {\sf postproc1} and {\sf postproc2} post-processing treatments of the classical model described in Sect.~\ref{subsection:postproc}. The dots on the black and orange tracks correspond to the evolutionary point at which profiles of $T$, $\rho$, and $r$ are shown in Fig.~\ref{fig:profiles_WR}. \textit{Right}: Kippenhahn diagram showing the evolution of the internal structure of our 60~$M_{\odot}$ stellar models. As the diagram is identical for both CMFGEN-based atmosphere and classical models, we only show it for one of them. Physical parameters of models corresponding to the points labeled with numbers in both plots are given in Table~\ref{tab:comparative_evolution_surf}.}
        \label{fig:hrd_WR}
    \end{figure*}

\begin{table*}[htp]
    \centering
    \caption{Evolution of the characteristic surface quantities of the classical model (1) and the CMFGEN-based model (2) for a 60~$M_\odot$ nonrotating solar metallicity star.}
    \label{tab:comparative_evolution_surf}
\begin{tiny}
\begin{tabular}{ccccccccccccccccccccccc}
\hline
\hline
{ID} &
      {$t$~[Myr]} &
      {$M_*$} &
      \multicolumn{2}{c}{$R_{\rm eff}$} &
      \multicolumn{2}{c}{$T_{\rm eff}$~[K]} &
      {$\Delta T_{\rm eff}$~[K]} &
     {$\log(L)$} &
      {$\dot{M}$~[$M_{\odot}$/yr]} &
     {$^{1}$H} &
    {$^{4}$He} &
     {$^{12}$C} &
      {$^{14}$N} &
      {$^{16}$O} \\
 & & & (1) & (2) & (1) & (2) & &  &  & & &  & &  \\
\hline
1 & 3.825 & 28.93 & 35.29 & 45.53 & 30$\,$601 & 26$\,$943 & 3$\,$658 & 5.991 & 5.32$\times 10^{-5}$ & 0.21 & 0.78 & 1.0$\times 10^{-4}$ & 8.3$\times 10^{-3}$ & 5.9$\times 10^{-5}$\\
2 & 3.888 & 26.34 & 4.50 & 7.52 & 81$\,$604 & 63$\,$122 & 18$\,$482 & 5.906 & 3.22$\times 10^{-5}$ & 0.21 & 0.78 & 6.3$\times 10^{-5}$ & 8.4$\times 10^{-3}$ & 6.0$\times 10^{-5}$\\
3 & 3.920 & 25.34 & 2.54 & 3.69 & 108$\,$206 & 89$\,$882 & 18$\,$324 & 5.901 & 2.45$\times 10^{-5}$ & 0.07 & 0.92 & 7.0$\times 10^{-5}$ & 8.4$\times 10^{-3}$ & 5.3$\times 10^{-5}$\\
4 & 3.940 & 24.91 & 2.22 & 3.97 & 115$\,$682 & 86$\,$445 & 29$\,$237 & 5.898 & 2.08$\times 10^{-5}$ & 0.00 & 0.99 & 1.1$\times 10^{-4}$ & 8.3$\times 10^{-3}$ & 4.4$\times 10^{-5}$\\
5 & 4.073 & 21.80 & 2.09 & 3.70 & 116$\,$867 & 87$\,$757 & 29$\,$110 & 5.863 & 2.61$\times 10^{-5}$ & 0.00 & 0.98 & 8.4$\times 10^{-3}$ & 8.1$\times 10^{-3}$ & 1.0$\times 10^{-3}$\\
6 & 4.127 & 20.32 & 1.52 & 4.54 & 136$\,$481 & 78$\,$932 & 57$\,$549 & 5.860 & 3.42$\times 10^{-5}$ & 0.00 & 0.30 & 0.52 & 2.7$\times 10^{-11}$ & 0.17\\
\hline \\
\end{tabular}
\tablefoot{Only one value is given for those parameters that remain unaffected by the change in the atmosphere treatment. $\Delta T_{\rm eff}$ refers to the effective temperature differences between model (1) and model (2). Surface abundances are given in mass fractions and the stellar mass, radius, and luminosity are given in solar units.}
\end{tiny}
\end{table*}

    Let us start with a general description of how models evolve in the HR diagram (see Fig.~\ref{fig:hrd_WR} and Table~\ref{tab:comparative_evolution_surf} for the definition of numbers). First the star evolves toward higher effective temperatures and lower luminosities (ID~1-2). When the hydrogen starts to be depleted at the surface because of the strong mass loss (ID~2), the star evolves at almost constant luminosity toward higher effective temperatures. The thinning of the envelope causes the hydrogen burning shell to shrink and eventually disappear (ID~3). When the hydrogen has totally disappeared from the surface of the star (ID~4), the star starts evolving at an almost constant effective temperature and toward lower luminosities. 
    The strong mass loss keeps on peeling off the outer layers of the star, while the convective core recedes in mass until shells previously processed by the 3-$\alpha$ nuclear reactions and enriched in carbon and oxygen are exposed at the surface (ID~5). Toward the end of this phase, the effective temperature increases again (ID~5,6), and the luminosity starts increasing again shortly before the end of core helium burning (ID~6).
    
    The effects of including detailed atmospheres on stellar evolution calculations are shown in Fig. \ref{fig:hrd_WR} and summarized in Table~\ref{tab:comparative_evolution_surf}. We find that at the start of the WR phase the CMFGEN-based model is close to the classical one in the Hertzsprung-Russell diagram. During its blueward evolution, the effective temperature of the CMFGEN-based model becomes increasingly lower than that of the classical one. The luminosity drop (ID~4-5), in which the models spend an important part of the evolution at relatively constant effective temperatures, is thus significantly shifted toward lower effective temperatures with differences larger than 25~kK (ID~4). The reversal of the hook at the end of our CMFGEN-based model is likely caused by numerical artifacts due to the grid configuration, and should not be compared to other tracks.

    In Table~\ref{tab:comparative_evolution_surf} we see that apart from the photospheric radius and effective temperature all the other physical properties remain unaffected by the change of the atmosphere treatment in the stellar model, as previously indicated by the internal temperature, density, and radius profiles. In particular, the models lose the same amount of mass as time proceeds whatever the treatment of the atmosphere because the mass loss prescription from \cite{Sander_Vink_2020} used in the computations only depends on the luminosity, mass, and surface hydrogen abundance, which all remain the same in both models. In the absence of rotational mixing, the surface abundances result directly from the mass loss, which is why they do not depend on the atmosphere treatment either. Consequently the internal structure, shown by the Kippenhahn diagram in the right panel of Fig.~\ref{fig:hrd_WR}, is unaffected by the treatment of the atmospheric boundary conditions.

    \subsection{Post-processing} \label{subsection:postproc}

    Considering that the feedback introduced in our CMFGEN-based models should be weak, we explored the possibility of recovering the results presented in the previous sections by applying post-processing methods.
    To do so we used the outputs of our classical models and applied two different post-processing treatments to them.
    
    \paragraph{Method postproc1} For each evolutionary point of the classical model, we retrieved the values of $T(\tau_{\rm match})$, $\log(g)$, and $\log (\dot{M})$ from the outputs and interpolated within the CMFGEN model atmospheres grid at this point. We obtained an interpolated radius profile that provided the associated photospheric radius, $R_{\rm eff}$, which in turn we combined with the evolutionary luminosity, $L_*$, from the classical model outputs to compute the corrected effective temperature, $T_{\rm eff,potsproc1}$. This is a similar experiment as the one done in \cite{Groh_et_al_2014}, but in our case it was done with interpolated model atmospheres and it was performed for all the time steps instead of only a select number of them. As is shown in Fig.~\ref{fig:hrd_WR}, this post-processing method predicts effective temperatures that are lower than the ones from the classical model by about 17~kK near ID~4 point, but higher than the ones from the CMFGEN-based model by about 12~kK at the same point in the evolution. \\

    \paragraph{Method postproc2} In this method we applied the same procedure as for Method {\sf postproc1} except for the definition of $\log(g)$ in the interpolation procedure. This time, instead of using the value from the classical evolutionary model output, we recomputed the surface gravity from the stellar mass, $M_*$, of the classical model and the photospheric radius, $R_{\rm eff}$, from the interpolated model at the previous time step. The resulting corrected effective temperature is noted as $T_{\rm eff,postptoc2}$.
    
    As is shown in Fig.~\ref{fig:hrd_WR}, this method reproduces fairly well the results obtained for our CMFGEN-based model. This confirms that the impact of the feedback from the atmosphere on the evolution model is small, which was already indicated by the fact that the classical model and the CMFGEN-based model have identical stellar parameters apart from their radius and effective temperature.
    The differences between our CMFGEN-based model and our {\sf postproc1} model are thus not due to the feedback of the atmosphere on the evolution model, but rather to the value of the gravity at the photosphere, $g$, which we used to interpolate our atmosphere profiles.

    \section{Comparisons with previous results} \label{section:comparison}

    In this section we now compare our CMFGEN-based models tracks to a selection of publicly available tracks (Sect.~\ref{s_comp_mod}). We also compare our results to observations of Galactic and LMC WR stars in Sect.~\ref{s_comp_ob}.

    \begin{figure}
        \centering
        \includegraphics[width=0.95\hsize]{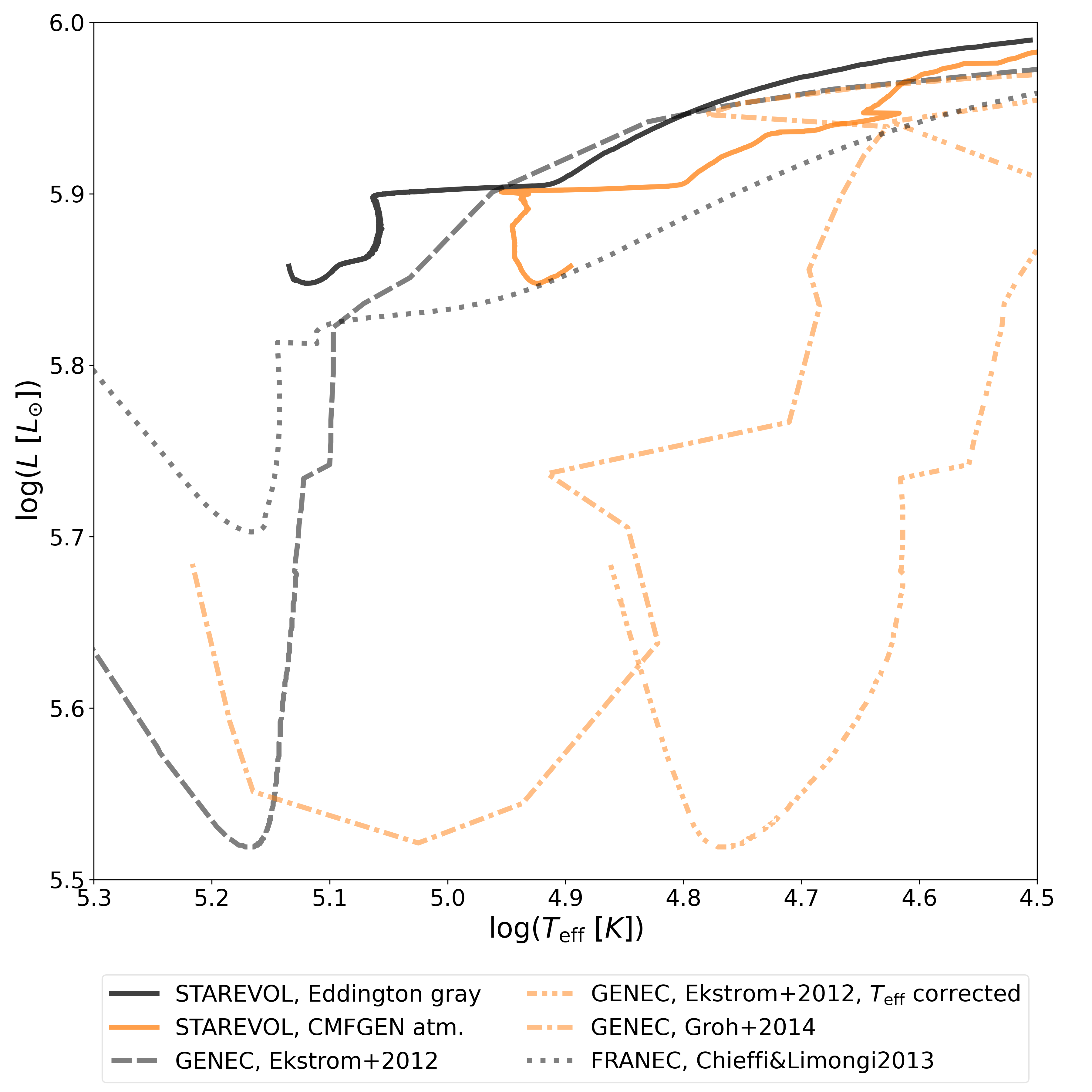}
        \caption{Evolutionary track of the nonrotating 60~$M_{\odot}$ stellar model at solar metallicity with an interpolated CMFGEN atmosphere included with feedback (solid thick orange line) in the HR diagram. The evolutionary track of a similar classical model (solid thick black line) is plotted for comparison. Additional nonrotating models are included: the classical model from \cite{Ekstrom_et_al_2012} without (dashed black line) and with (dashed double-dotted orange line) the effective temperature correction from \cite{Schaller_et_al_1992}; the \cite{Groh_et_al_2014} model (dash-dotted orange line) using post-processed temperatures from CMFGEN calculations; and the classical \cite{Chieffi_Limongi_2013} model (dotted black~line).}
        \label{fig:hrd_with_tracks}
    \end{figure}
    
    \begin{figure}
        \centering
        \includegraphics[width=0.95\hsize]{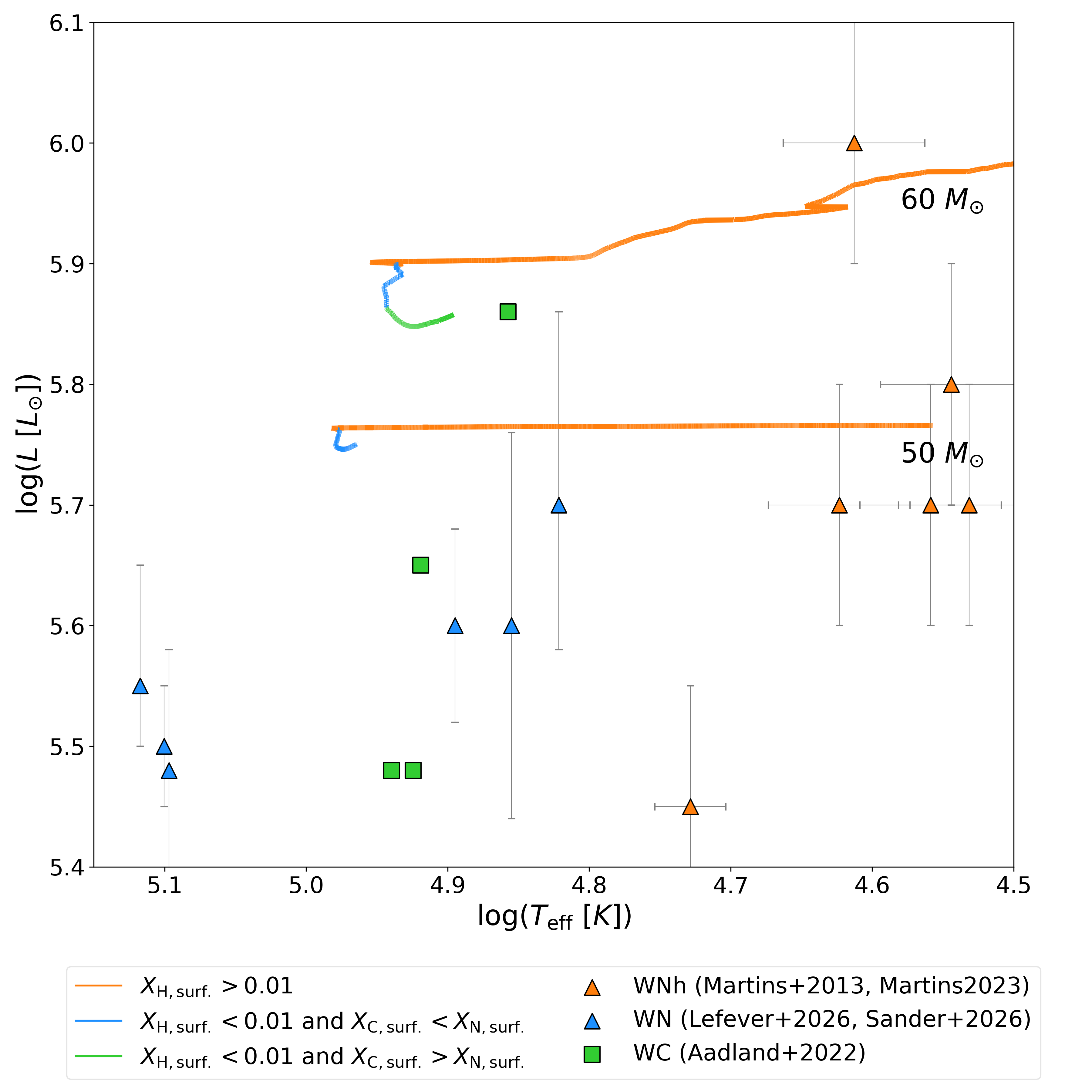}
        \caption{Evolutionary tracks of our nonrotating stellar models with initial masses of 50~$M_{\odot}$ and 60~$M_{\odot}$ at solar metallicity, taking into account detailed model atmosphere with feedback. Observations of galactic WN stars from \cite{Martins_et_al_2013}, \cite{Martins_2023}, and \cite{Lefever_et_al_2026}, of WN2 stars in the galaxy and the LMC from \cite{Sander_et_al_2026} and of WC stars in the LMC from \cite{Aadland_et_al_2022}, are plotted for comparison. The evolutionary tracks and the observation points are color-coded to show the different spectral types (WNh, hydrogen-free WN, and WC).}
        \label{fig:hrd_with_obs}
    \end{figure}

    \subsection{Comparison with models}
    \label{s_comp_mod}

    We show in Fig.~\ref{fig:hrd_with_tracks} the evolutionary tracks of our 60~$M_{\odot}$ models in the HR diagram compared to evolutionary tracks from nonrotating models with the same mass and metallicity from the literature. 
    All the classical evolution models \citep[i.e. those of][and our own]{Ekstrom_et_al_2012,Chieffi_Limongi_2013} predict effective temperatures exceeding 100~kK. They show drops in luminosity with different amplitudes as the mass loss prescriptions used are different. However, they all spend the majority of their time on the WR phase on this luminosity drop, which is located in a region of very high effective temperatures. 
    
    We see that including more detailed model atmospheres significantly reduces the effective temperature of the evolution models when compared to the classical case. The model from \cite{Groh_et_al_2014}, which includes temperature corrections during post-processing based on detailed model atmospheres computed with CMFGEN, predicts an effective temperature shift compared to the classical model that is similar to the one we obtain. The GENEC model from \cite{Ekstrom_et_al_2012} with the post-processing correction from \cite{Schaller_et_al_1992} overestimates this effective temperature shift, as was already shown by \cite{Groh_et_al_2014}.

    \subsection{Comparison with observations}
    \label{s_comp_ob}

    Figure~\ref{fig:hrd_with_obs} shows our 50 and 60~$M_{\odot}$ CMFGEN-based evolution models along with observations of WR stars. Our models allow one to define an effective temperature that truly corresponds to a photospheric value, at $\tau =2/3$. This is different from the majority of publicly available tracks, for which the temperature quoted in evolutionary calculations better corresponds to an optical depth larger than $\sim$10, and is usually referred to as $T_*$. For the evolutionary phases with relatively weak winds, the difference between $T_{\rm eff}$ and $T_*$ is negligible, but for phases with dense winds it can amount to several thousands Kelvins. As a consequence, it is common practice to quote only $T_*$ for stars analyzed by atmosphere models, in order to compare their surface properties with evolutionary models \citep[e.g.][]{Hamann_et_al_2006,Hamann_et_al_2019,Sander_et_al_2019}.
    We thus include in Fig.~\ref{fig:hrd_with_obs} objects for which effective temperatures, given at an optical depth equal to 2/3, are available in the literature. Galactic WN stars studied by \cite{Martins_et_al_2013}, \cite{Martins_2023}, \cite{Sander_et_al_2026}, and \cite{Lefever_et_al_2026} satisfy this criterion. Since there were so few objects, we also added a few WC4 and WN2 stars located in the LMC \citep{Aadland_et_al_2022,Sander_et_al_2026}. Given that we adopt relatively low mass loss rates in our computations (see Sects.~\ref{section:tools} and~\ref{s_mdot}) and that mass loss rates are lower at lower metallicity \citep[e.g.][]{Crowther_et_al_2002}, including LMC objects is reasonable.
    
    For a finer comparison of our models with these objects, we assume the following relations between chemical surface composition and spectral types: $X_{\rm H, surf.} > 0.01$ corresponds to WNh stars; and $^{12}$C/$^{14}$N $\ll$ ($\gg$) 1 corresponds to WN (WC) stars \citep[see][]{Meynet_Maeder_2003}.

    Our models are fully able to reproduce the position of WNh stars (orange lines and symbols in Fig.~\ref{fig:hrd_with_obs}). The models transition from WNh to H-free WN stars at $T_{\rm eff} \sim$ 90$\,$000~K, slightly above the position of the three WN stars studied by \citet{Lefever_et_al_2026}. This is still a major improvement compared to classical models that never exhaust their surface hydrogen below 10$^5$~K (see Fig.~\ref{fig:hrd_with_tracks}). We note that the three WN2 stars of \citet{Sander_et_al_2026} remain out of reach of our models.
    Regarding WC stars the carbon-rich phase of our 60~$M_\odot$ track takes place at effective temperatures close to 85$\,$000~K, in excellent agreement with the temperature of the WC4 stars shown in Fig.~\ref{fig:hrd_with_obs}. In that case too, literature models overestimate the observed effective temperatures. Our 50~$M_{\odot}$ evolution model does not become a WC star as the mass loss rates we used are too low to expose the carbon-rich core of the star and we do not take into account extra mixing due to rotation. This will be adjusted in future studies.

    While these comparisons remain limited by the small number of observations available, they still show that a proper treatment of the atmosphere in evolutionary calculations allows for a significant improvement in the understanding of WR stars' evolution and surface properties.

\section{Limitations} \label{section:limitations}

    In the previous sections, we presented the models we obtained using the method we developed to include detailed atmospheres in evolution computations. In this section we present some limitations of these models.

   \subsection{Mass loss} \label{s_mdot}

    \begin{figure}
        \centering
        \includegraphics[width=0.95\hsize]{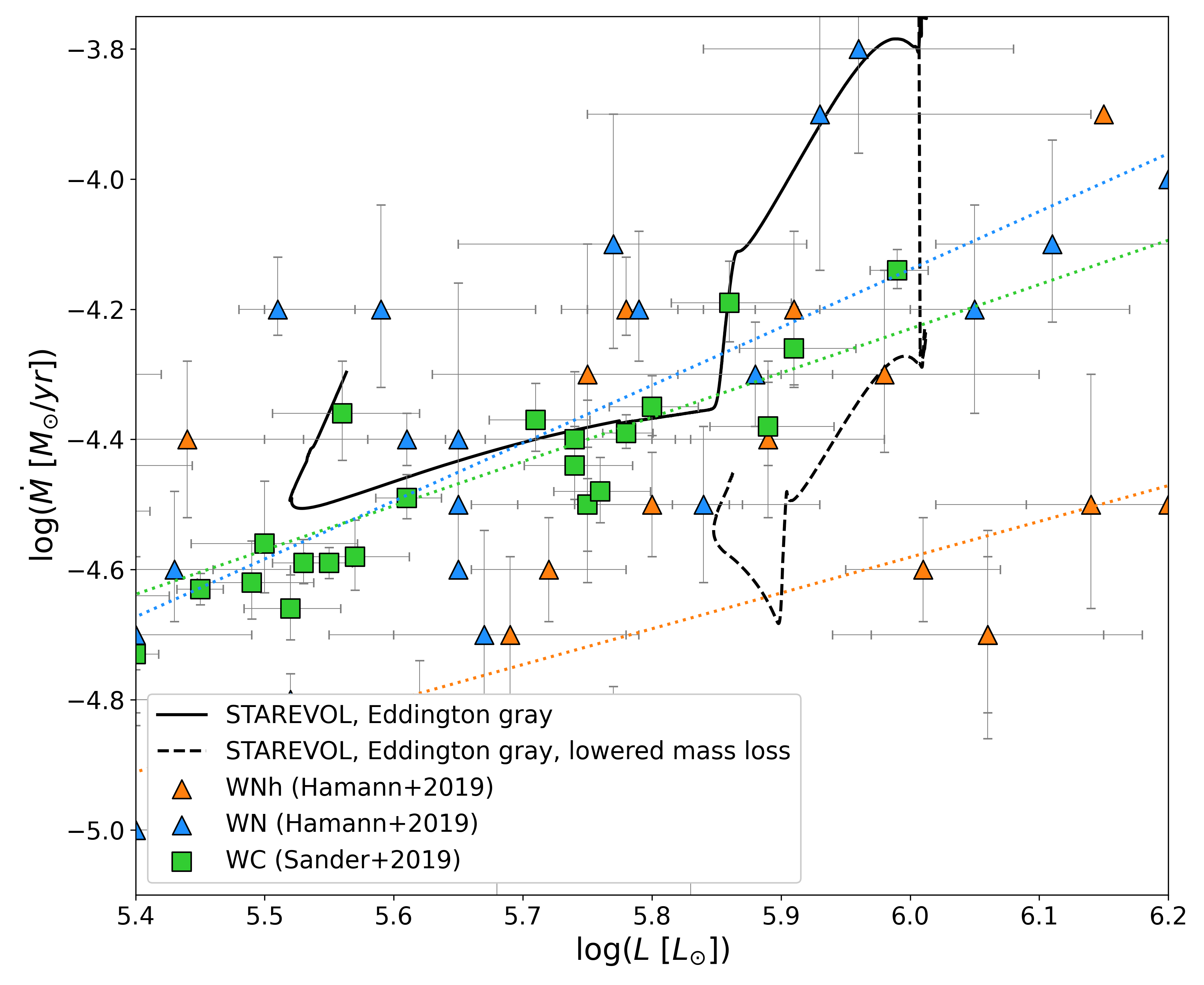}
        \caption{Mass loss as a function of luminosity for a 60~$M_{\odot}$ classical nonrotating model without (solid black line) and with (dashed black line) the reduction in the mass loss rate by 0.5 dex presented in Sect.~\ref{subsection:starevol_description}. Observations of WN stars from \citet[orange and blue triangles]{Hamann_et_al_2019} and WC stars from \citet[green squares]{Sander_et_al_2019} are also presented along with the mass loss relations derived in their papers (dotted lines of the corresponding colors).}
        \label{fig:mlos_L_with_obs_and_tracks}
    \end{figure}

    As explained in Sect.~\ref{subsection:starevol_description}, we used a mass loss rate prescription developed for hydrogen-free stars, which we also extended to the WNh phase. However, contrarily to the hydrogen-free WR phases, the wind in the WNh phase is not launched from the iron-opacity bump \citep{Sander_et_al_2023}. The impact of the hydrostatic radius shown in \cite{Sander_et_al_2023} was also not taken into account, which might impact the strength of the wind before or after the luminosity drop. This prescription was also reduced by 0.5 dex in order to facilitate the computation of the CMFGEN model atmospheres. 
    
    We show in Fig.~\ref{fig:mlos_L_with_obs_and_tracks} two classical 60 $M_\odot$ evolution models, one computed with the reduced mass loss rates and one computed with the non-reduced mass loss rates, along with observation points and empirical linear $\dot{M}$ - $L$ relations from \cite{Hamann_et_al_2019} and \cite{Sander_et_al_2019}. Our model with non-reduced mass loss rates vastly overestimates the mass loss rates of WNh stars. However, it is more consistent with observational mass loss rates of hydrogen-free WN stars and of WC stars. Our model with reduced mass loss rates is consistent with the mass loss rates of WNh stars and underestimates the ones of hydrogen-free WN stars and of WC stars, although it remains close to the observational error bars. The mass loss rates we adopted can thus be seen as a compromise between the different mass loss rates observed for different WR regimes.

    These reduced mass loss rates affect the evolution of the luminosity of the models and reduce their luminosity span on the WR phase compared to models with non-reduced mass loss rates. 
    In our method we interpolated model atmosphere profiles whose temperature structure matched the one of the evolution model in the inner regions of the atmosphere. This choice implies that the gravity at the photosphere and the mass loss rate impact the effective temperature of the interpolated model, and thus its photospheric radius. The results of our method are thus dependent on the mass loss rate prescription used in the stellar evolution model. Because of the increase in the atmosphere extension due to the stellar wind, higher mass loss rates are expected to bring about higher photospheric radii, and thus lower effective temperatures. As we chose to use lower mass loss rates, the differential effect we show is a lower limit of what is expected for the hydrogen-free WR phases.

    \subsection{The inner atmosphere} \label{subsection:inner_atmosphere}
 
    At the beginning of the WR phase, the inner region of the atmosphere of the models corresponds to the iron opacity bump \citep{Cantiello_et_al_2009}. 
    This opacity bump lowers the local Eddington luminosity, $L_{\rm Edd} = 4 \pi c G M / \kappa$, which can become lower than the local radiative luminosity, $L_{\rm rad}$. In 1D evolution codes, this can trigger convection and/or a density inversion \citep{MESA_2_Paxton_et_al_2013}. This subsurface convection zone is inefficient in terms of energy transportation. As said before this leads to numerical problems in 1D stellar evolution codes, which can render the computation of the model very slow and difficult or even stop the computation altogether \citep{Agrawal_et_al_2022}. 
    
    The iron opacity bump, along with other opacity peaks which appear at lower temperatures, can also trigger envelope inflation \citep{Ishii_et_al_1999,Grafener_et_al_2012,Sanyal_et_al_2015} in the evolution model. The density is thus lowered in the regions associated with the opacity peaks.
    This effect is particularly important in the cold phases of the evolution, where the opacity peaks are located at high optical depth, and thus affect the radius profile in the region where we patched the models.
    We therefore underline the necessity of a better understanding of these subsurface regions, along with the inclusion of the related physics in state-of-the-art model atmospheres and evolution codes \citep[see for instance][]{Moens_et_al_2022,Debnath_et_al_2024}. A temporary solution was suggested recently by \cite{Josiek_et_al_2025}, who studied the impact of the point at which model atmospheres are patched in the post-processing correction of a 150~$M_{\odot}$ evolution model. They showed that patching the model atmospheres deeper in the star below the iron opacity bump would allow us to avoid dealing with inconsistencies between evolution models and model atmospheres.

\section{Conclusion} \label{section:conclusion}

    In this study we tested the differential impact of including detailed atmospheres in stellar evolution models of WR stars when compared to the classical Eddington gray atmosphere treatment. We presented a method of including state-of-the-art model atmospheres in massive stars evolution computations. This method relies on a precomputed grid of detailed model atmospheres. In the process of solving the stellar structure and evolution equations, the surface boundary conditions to the mass conservation and energy transport equations are interpolated within the grid of atmosphere models and applied at an optical depth large enough to ensure the diffusion approximation to be valid. The associated atmosphere profiles then provide the thermal and density structure of the model at lower optical depth and are used to determine the photospheric radius and the effective temperature.
    
    We applied this method to the advanced phases of the evolution of nonrotating models at solar metallicity with initial masses of 50 $M_{\odot}$ and 60 $M_{\odot}$. We compared it to otherwise identical classical computations. We find that the differences in the effective temperature 
    are very significant, as the CMFGEN-based models predict effective temperatures lower than the classical ones by more than 20~kK for about half of the evolution time on the WR phase. This is mainly caused by the extension of the wind, which is not accounted for in the Eddington gray approximation and induces an important increase in the photospheric radius. As the mass loss rates we used are smaller than the ones classically used for WR stars, the differential impact we show on our model should be considered as a lower estimate. The reduction in the effective temperature allows us to better reproduce the position of Galactic and LMC WR stars in the HR diagram.

    We also showed that using detailed atmospheres has no impact on the internal structure and evolution of the model. It is only its appearance, in terms of effective temperature, that is affected. 
    We then presented a method to correct the radius and effective temperature of a precomputed classical evolution model in post-processing using detailed model atmospheres. Post-processing corrections that account for the modified effective radius and surface gravity lead to evolutionary tracks that are almost undistinguishable from the ones obtained with the more sophisticated method we developed.

    For the future we emphasize the need to test this method on a larger range of masses and metallicities to see its impact on evolution models, using higher mass loss rates in the hydrogen-free WN phase and the subsequent phases of the evolution in better agreement with observations.
    However, as the computational cost of the model atmosphere grid is quite high, thought and effort must be put into improving the efficiency of the computation of model atmosphere grids for massive stars. Efforts should also be put into solving the differences between the evolution and atmosphere models, and into correctly modeling the subsurface regions of massive stars. This would allow for a proper self-consistent treatment of the atmosphere for the entire evolution of massive stars, including the very poorly constrained cool phases of their evolution.

\begin{acknowledgements}
We thank the referee (A. Sander) for a prompt and positive report. We thank John Hillier for developing CMFGEN and making it available to the community.
\end{acknowledgements}

\bibliographystyle{aa}
\bibliography{bib}

\begin{appendix}

\onecolumn

\section{Grid of atmosphere models for the 50~$M_{\odot}$ track}
\label{ap_1}

\begin{figure*}[h]
        \centering
        \includegraphics[width=0.325\hsize]{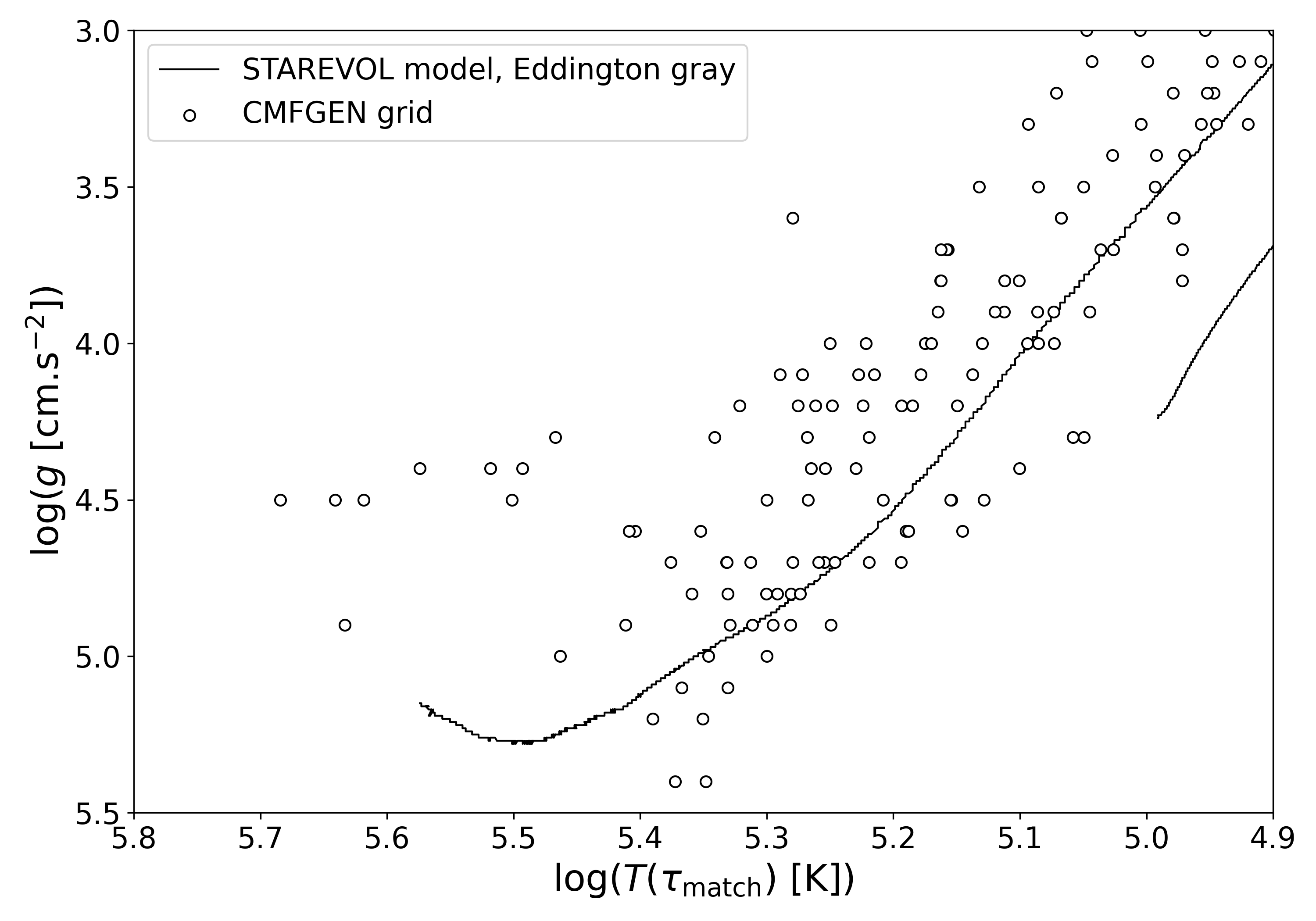}
        \includegraphics[width=0.325\hsize]{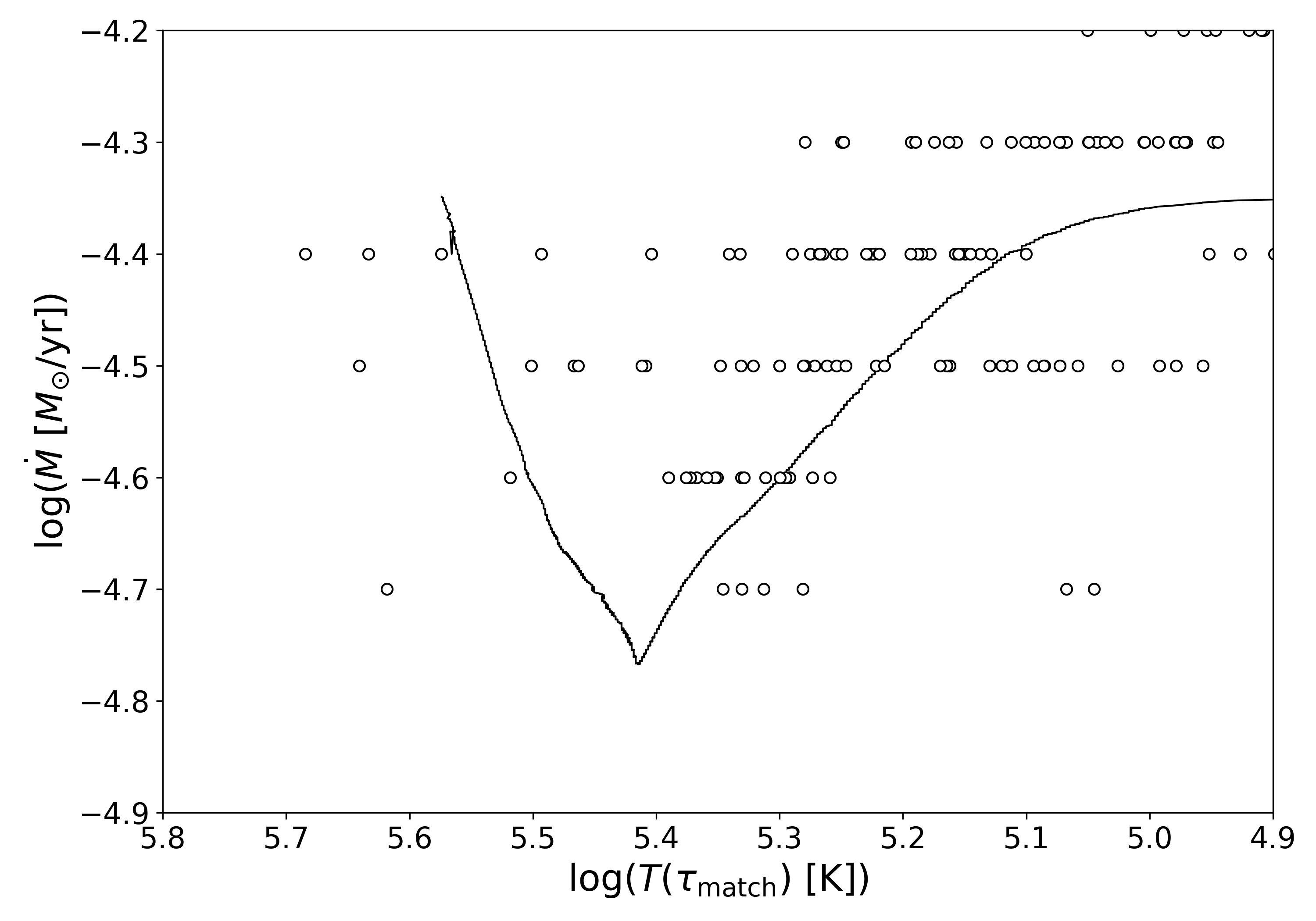}
        \includegraphics[width=0.325\hsize]{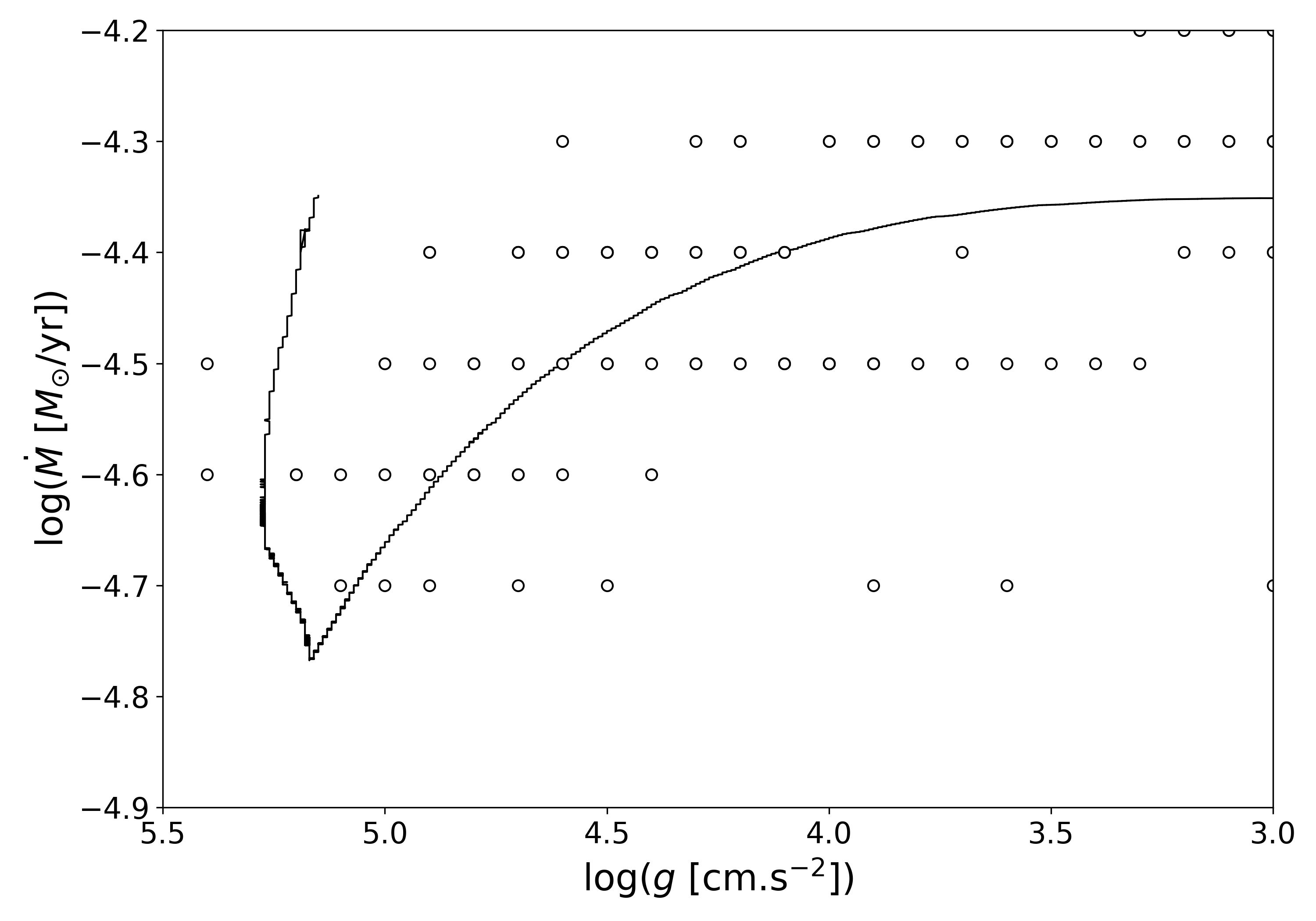}
        \caption{Same as Fig~\ref{fig:grid_60Msun} but for the 50~$M_{\odot}$ Eddington gray stellar model we used and the associated grid.}
        \label{fig:grid_50Msun}
    \end{figure*}

Figure~\ref{fig:grid_50Msun} shows the grid of CMFGEN atmosphere models we used to compute the 50~$M_{\odot}$ track with detailed atmospheres. 

\begin{table*}[h]
    \centering
    \caption{Evolution of the characteristic surface quantities of the classical model (1) and the CMFGEN-based model (2) for a 50~$M_{\odot}$ nonrotating solar metallicity star.}
    \label{tab:comparative_evolution_surf_2}
\begin{tiny}
\begin{tabular}{ccccccccccccccccccccccc}
\hline
\hline
{ID} &
      {$t$~[Myr]} &
      {$M_*$} &
      \multicolumn{2}{c}{$R_{\rm eff}$} &
      \multicolumn{2}{c}{$T_{\rm eff}$~[K]} &
      {$\Delta T_{\rm eff}$~[K]} &
     {$\log(L)$} &
      {$\dot{M}$~[$M_{\odot}$/yr]} &
     {$^{1}$H} &
    {$^{4}$He} &
     {$^{12}$C} &
      {$^{14}$N} &
      {$^{16}$O} \\
 & & & (1) & (2) & (1) & (2) & &  &  & & &  & &  \\
\hline
1 & 4.407 & 19.74 & 16.78 & 19.58 & 38$\,$974 & 36$\,$078 & 2$\,$896 & 5.766 & 4.44$\times 10^{-5}$ & 0.39 & 0.60 & 5.5$\times 10^{-5}$ & 8.0$\times 10^{-3}$ & 5.1$\times 10^{-4}$\\
3 & 4.417 & 19.38 & 2.72 & 4.71 & 96$\,$737 & 73$\,$539 & 23$\,$198 & 5.765 & 2.56$\times 10^{-5}$ & 0.15 & 0.84 & 6.1$\times 10^{-5}$ & 8.4$\times 10^{-3}$ & 6.3$\times 10^{-5}$\\
4 & 4.433 & 19.07 & 1.88 & 2.82 & 116$\,$156 & 94$\,$774 & 21$\,$382 & 5.761 & 1.71$\times 10^{-5}$ & 0.00 & 0.99 & 1.0$\times 10^{-4}$ & 8.3$\times 10^{-3}$ & 4.8$\times 10^{-5}$\\
6 & 4.517 & 17.52 & 1.61 & 2.95 & 124$\,$843 & 92$\,$117 & 32$\,$726 & 5.751 & 2.19$\times 10^{-5}$ & 0.00 & 0.99 & 1.4$\times 10^{-4}$ & 8.3$\times 10^{-3}$ & 4.6$\times 10^{-5}$\\
\hline \\
\end{tabular}
\tablefoot{The ID~2 point (start of the surface hydrogen depletion) is located before we start to include detailed atmospheres in our 50~$M_{\odot}$ model. As described in Sect.~\ref{section:comparison} the ID~5 point ($^{12}$C/$^{14}$N $\ge 1$) never happens in our 50~$M_{\odot}$ model.}
\end{tiny}
\end{table*}

\FloatBarrier

\end{appendix}

\end{document}